\documentclass[aps,prb,twocolumn,showpacs,superscriptaddress,amsmath,amssymb,floatfix,longbibliography]{revtex4-2}

\usepackage[breaklinks=true,colorlinks,citecolor=blue,linkcolor=blue,urlcolor=blue]{hyperref}
\usepackage{float}
\usepackage{epsfig,mathrsfs,color,latexsym,subfigure,marginnote}
\usepackage{graphicx,physics,longtable,amsmath,amssymb,amsfonts}
\usepackage{multirow}

\usepackage{ulem}
\usepackage{array}

\begin{document}

\title{Role of effective mass anisotropy in realizing a hybrid nodal-line fermion state}

\author{Bikash Patra}
\thanks{These authors contributed equally to this work.}
\affiliation{Department of Condensed Matter Physics and Materials Science, Tata Institute of Fundamental Research, Mumbai 400005, India}

\author{Rahul Verma}
\thanks{These authors contributed equally to this work.}
\affiliation{Department of Condensed Matter Physics and Materials Science, Tata Institute of Fundamental Research, Mumbai 400005, India}

\author{Shin-Ming Huang}
\affiliation{Department of Physics, National Sun Yat-sen University, Kaohsiung 80424, Taiwan}
\affiliation{Center for Theoretical and Computational Physics, National Sun Yat-sen University, Kaohsiung 80424, Taiwan}
\affiliation{Physics Division, National Center for Theoretical Sciences, Taipei 10617, Taiwan}

\author{Bahadur Singh}
\email[Corresponding author:~]{bahadur.singh@tifr.res.in}
\affiliation{Department of Condensed Matter Physics and Materials Science, Tata Institute of Fundamental Research, Mumbai 400005, India}


\begin{abstract}
Understanding the role of lattice geometry in shaping topological states and their properties is of fundamental importance to condensed matter and device physics. Here we demonstrate how an anisotropic crystal lattice drives a topological hybrid nodal line in transition metal tetraphosphides $M$P$_4$ ($M$ = Transition metal). $M$P$_4$ constitutes a unique class of black phosphorus materials formed by intercalating transition metal ions between the phosphorus layers without destroying the characteristic anisotropic band structure of the black phosphorous. Based on the first-principles calculations and $k \cdot p$ theory, we show that $M$P$_4$ harbor a single hybrid nodal line formed between anisotropic $M~d$ and P states with oppositely-oriented effective masses unhinged from the high-symmetry planes. The nodal line consists of both type-I and type-II nodal band crossings whose nature and location are determined by the effective mass anisotropies of the intersecting bands. We further discuss a possible topological phase transition to exemplify the formation of the hybrid nodal line state in $M$P$_4$. Our results offer a comprehensive study for understanding the interplay between structural motifs-driven mass anisotropies and topology in anisotropic lattice materials to realize hybrid semimetal states.
\end{abstract}

\maketitle
\clearpage

\section {Introduction} 
Since the discovery of topological insulators, the topology of electronic states and finding robust topological materials have attracted broad interest~\cite{hasan2010colloquium,bansil2016colloquium, RMP_WeylAshvin,singh2022topology}. The initial topological state characterization based on free-fermion symmetries such as time-reversal was generalized later to include topological states protected by crystalline symmetries~\cite{fu2011topological,hsieh2012topological, RTCI_Fu2019, Multi_Fermion}. In this way, many topological crystalline states such as mirror, inversion, or rotational Chern insulators, topological semimetals with Dirac, Weyl, nodal-line, nodal-chain, and higher-fold chiral fermions, among other possibilities, were proposed with their unique set of nontrivial states and electromagnetic properties~\cite{bansil2016colloquium, RMP_WeylAshvin,singh2022topology,fu2011topological,hsieh2012topological, RTCI_Fu2019, Multi_Fermion, Berryphase_Xiao}. The topological states in materials are described uniquely by topological invariants that depend only on the symmetries of their underlying atomic positions~\cite{Slager2013,kruthoff2017topo,TQC_Bradlyn2017, SI_Song2018}. Such topological state characterization has allowed unique symmetry-to-topology mapping that facilitated the identification of topological states in high-throughput materials searches, revealing many topological materials with unique nontrivial states~\cite{Database1_CFang, Database2_AB2019, Database3_Ashvin2019}. These topological searches often neglect the effects of the spatial arrangement of atoms and their wavefunction properties even though they are essential for describing the numbers, energy-momentum relations, and geometries of the nontrivial states~\cite{Dandling_Lin2013, Singh_termination,singh2018saddle}.

Among various topological states, topological nodal-line semimetals are considered parent phases for realizing many exotic topological gapped or ungapped states through systematic lowering of crystalline symmetries~\cite{NodalLine_Balents,Nodal_Liang,bian2016drumhead,chan2016ca,StarFruit_SM}. They form one-dimensional (1D) valence and conduction band crossings in the momentum space, which realize new quasiparticles beyond the well-known Dirac, Weyl, and Majorana particles in high-energy physics. Analogous to Dirac and Weyl point semimetals, topological nodal-line semimetals can be classified as type-I, type-II, or hybrid nodal line semimetals depending on the fermiology of the crossings bands~\cite{TypeII_LaAlGe,chang2019realization,wang2019topo,hybrid_Weyl,hybrid_zhang2018}. In the type-I case, a nodal line is formed between electronlike and holelike bands and respects Lorentz symmetry. In type-II semimetals, the nodal line is formed between either two electronlike or holelike bands, strongly breaking the Lorentz symmetry. In contrast, the hybrid nodal line consists of both the type-I and type-II band degeneracies that are naturally expected to occur if one of the crossing bands has a saddlelike energy dispersion. Such band crossings constitute a Fermi surface with electron and hole pockets that touch along specific directions in $k$ space. The hybrid nodal line can realize amplified electron-correlation effects owing to the saddlelike energy dispersion and directional-dependent magnetotransport properties due to the Klein tunneling between the electron and hole pockets, among other phenomena~\cite{NodalLine_Balents,hybrid_zhang2018,hybrid_Weyl,Correlation_Shao2020}.

\begin{figure}[h!]
\centering
\includegraphics[width=0.5\textwidth]{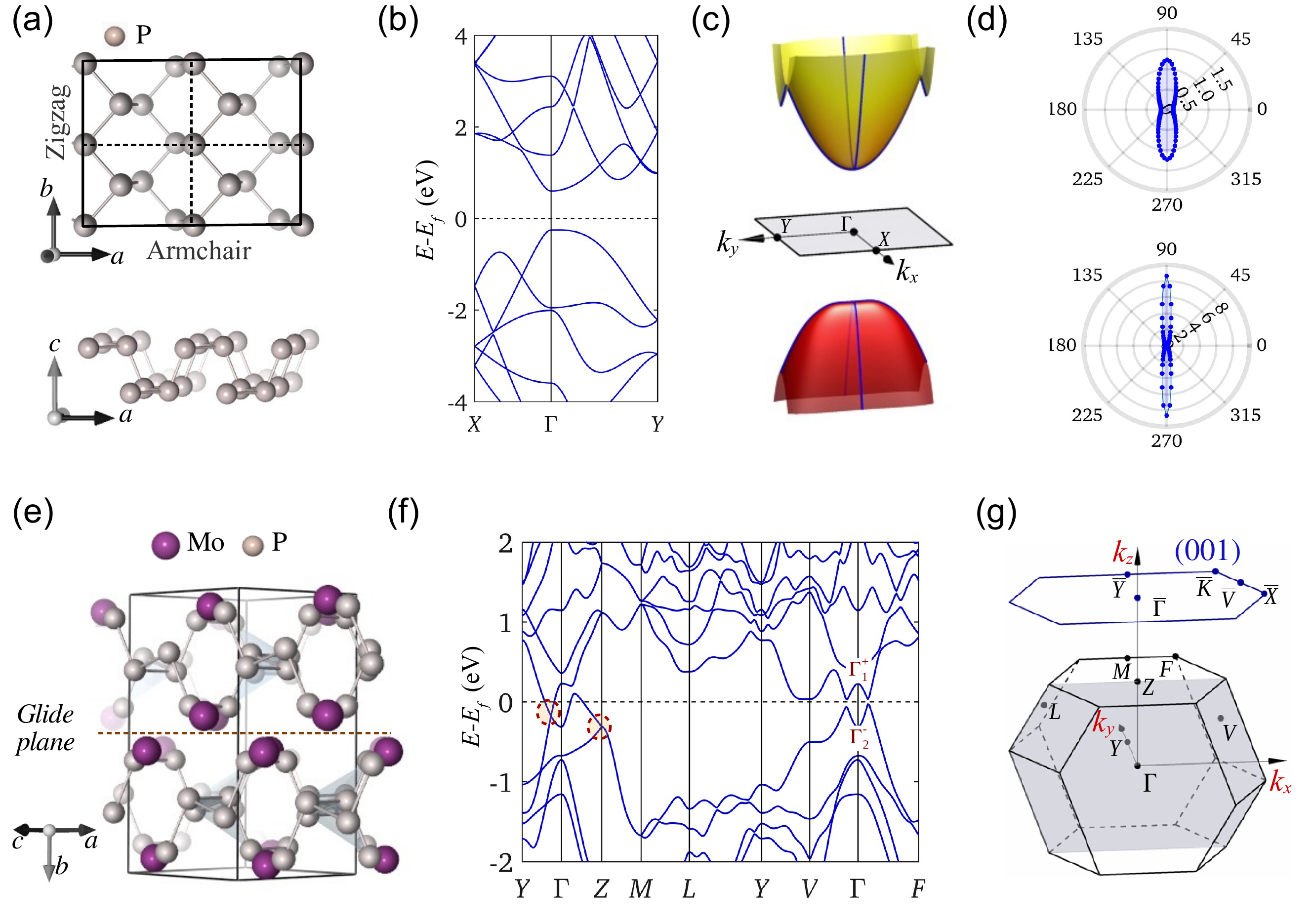}
\caption{{\bf Crystal lattice and electronic structure of phosphorene and MoP$_4$.}  (a) Puckered honeycomb structure of a single phosphorene layer with distinct P-P bonding arrangements along the armchair ($x$ axis) and zigzag ($y$ axis) directions. The in-plane bonding anisotropy is the origin of the directional selective properties of phosphorene. (b)-(c) Band structure of a single phosphorene layer along the high-symmetry directions (b) and in the full two-dimensional Brillouin zone (c). (d) Calculated electron (top) and hole (bottom) carrier effective masses (in the units of mass of electron $m_e$) of a single phosphorene layer in polar coordinates. $\theta =0^{\circ}$ and $\theta =90^{\circ}$ mark the armchair and zigzag directions as shown in (a). (e) Perspective view of the bulk crystal lattice of MoP$_4$ with $C2/c$ (No. 15) symmetry. The dashed horizontal line marks the glide mirror plane. (f) First-principles semimetallic band structure of MoP$_4$ without spin-orbit coupling. Irreducible representations at $\Gamma$ point are marked. The symmetry-protected band crossings near the Fermi level are highlighted in dashed red circles. (g) The primitive cell Brillouin zone of MoP$_4$ with high-symmetry points. The mirror plane is in highlighted grey, and the projected (001) surface Brillouin zone in blue is shown.}\label{Fig_CES}
\end{figure}

Because the hybrid nodal lines constitute both the type-I and type-II band crossings located along different momentum space directions, their experimental realization remains a daunting challenge. Here we propose that effective masses of the intersecting bands in anisotropic lattice materials can determine the dispersion and location of the nodal band crossings.  Of importance is orthorhombic black phosphorus in which individual two-dimensional (2D) phosphorene layers are stacked together by weak Van der Waals interactions~\cite{BP_Review_Kou2015,xia2019black,BP_Qiao2014,BP_Xia2014,xia2019black,SPR_2DXinming2017}. Inside each phosphorene layer, the phosphorus atoms form puckered honeycomb arrangement along the armchair and a bilayer configuration in the zigzag directions [see Fig.~\ref{Fig_CES}(a)-(d)]. This unique structural motif drives the anisotropic electronic structure and physical properties that facilitate the use of black phosphorus in devices with directional selectivity~\cite{BP_Review_Kou2015,xia2019black,BP_Qiao2014,BP_Xia2014,xia2019black}. The electronic structure and physical properties of black phosphorus can be engineered by pressure, doping, adsorption, and electric-field controls to realize insulator-to-metal transition, topological semimetals with anisotropic Dirac bands, and superconductivity without destroying their anisotropic character~\cite{xiang2015pressure,Phosphorene_DSM,Phosphorene_Efield,ghosh2016electric,chris2022superconductivity}. Importantly, the intercalation of transition metal atoms in black phosphorus generates varied low-symmetry transition metal tetraphosphides $M$P$_4$ ($M$ = Transition metal) that preserve the anisotropic electronic bands of black phosphorus but with a semimetallic ground state~\cite{TMBP_Gong2018,BPNa_Sun2015,cheng2017sodium,TmP4_CS,MoP4_HighPressure,TmP4_CS_WP4,TmP4_Khan2020}. Here, we show that the $M$P$_4$ realizes a single hybrid nodal line driven by mass-anisotropies of $M$$~d$ and P bands. By carefully exploring the structure-property-topology mapping in $M$P$_4$, our results demonstrate how an anisotropic crystal structure shapes the low-energy topological states in materials.

\section{Methods}
Electronic structure calculations were performed within the density functional theory (DFT) framework with the projector augmented wave~\cite{DFT_1964,blochl1994projector} potentials as implemented in the Vienna {\it ab-initio} simulation package (VASP)~\cite{kresse1996efficient,kresse1999from}. We used Perdew-Burke-Ernzerhof parameterized generalized gradient approximation (GGA)~\cite{perdew1996generalized} and hybrid Heyd-Scuseria-Ernzerof (HSE) functionals for the exchange-correlation effects. The spin-orbit coupling (SOC) was included self-consistently to incorporate relativistic effects in the calculations. An energy cut-off of 400 eV for the plane wave basis set and a $8\times 8 \times 10$ $\Gamma$-centered $k$ mesh for the Brillouin zone sampling were used. The experimental lattice parameters with fully relaxed atomic positions were adopted to calculate the electronic structure and topological properties of $M$P$_4$. We constructed material-specific tight-binding model Hamiltonian from the atoms-centered Wannier functions~\cite{arash2014an}. $M$ $d$ and P $s$ and $p$ orbitals were used to construct the Wannier functions. The topological properties and surface states calculations were performed using the WannierTools package~\cite{wu2018wanniertools}.

\section{Results}

\subsection{Crystal structure and anisotropic electronic properties}

The anisotropic properties of black phosphorus are incepted in the anisotropic lattice of phosphorene, which is a two-dimensional (2D) building block of black phosphorus~\cite{BP_Review_Kou2015,xia2019black,BP_Qiao2014,BP_Xia2014}. The P atoms in the phosphorene are covalently bonded with their three neighboring P atoms to form two in-plane and one out-of-plane bond, generating a hexagonal puckered lattice (Fig.~\ref{Fig_CES}(a)). This atomic arrangement drives a highly anisotropic energy dispersion such that the conduction and valence bands are nearly flat along the zigzag direction ($\Gamma-Y$) and significantly dispersive along the armchair direction ($\Gamma-X$) near the band extremum point (see Figs.~\ref{Fig_CES}(b)-(c)). Based on the energy-dispersion, $E = \frac{\hbar^2 k^2}{2m^*}$, where $m^*$ is the effective mass, we calculate the direction-dependent $m^*$ and show results in Fig~\ref{Fig_CES}(d). The effective masses are higher along the zigzag direction than the armchair direction for both the valence (hole) and conduction (electron) bands. The hole-effective masses are $8.62 m_e$ and $0.14 m_e$ ($m_e$ is the mass of an electron in vacuum) along the zigzag and armchair directions that yield a mass anisotropy of $\sim60$. Band crossings in such anisotropic bands can realize hybrid nodal dispersion under appropriate material parameters and symmetries as discussed below. 

Using phosphorene anisotropic band structure as a guideline, we searched among the black phosphorus materials and identified the most robust and ideal hybrid nodal-line candidate in transition metal tetraphosphides $M$P$_4$~\cite{MoP4_HighPressure,TmP4_CS,TMBP_Gong2018}. We discuss the structural and electronic properties of $M$P$_4$ by taking MoP$_4$ as an exemplary system. The single crystals of MoP$_4$ with a black-phosphorus-derived structure have been grown, and the transport experiments have reported it as a semimetal with large positive magnetoresistance~\cite{TmP4_CS,MoP4_HighPressure}. MoP$_4$ crystallizes in the monoclinic Bravais lattice with space group $C2/c$ (No. 15) that has lower symmetry than orthorhombic black-phosphorus  $Cmce$ (No. 64). The experimental structural parameters are $a=5.3131$ \r{AA}, $b = 11.1588$ \r{AA}, $c = 5.8343$ \r{AA}, and $\beta= 110.638^\circ$~\cite{MoP4_HighPressure}. The crystal structure is derived by intercalation of the Mo atoms in between the phosphorene layers, which reorders the atomic-stacking of phosphorene layers from AB to AC in a way similar to the alkali metals intercalation in black phosphorus~\cite{BPNa_Sun2015,cheng2017sodium}. Figure~\ref{Fig_CES}(e) depicts the unit cell of MoP$_4$. It consists of two phosphorene layers with Mo atoms sandwiched between them. The sandwiched Mo atoms form zigzag chains extending along [001] with a uniform interatomic distance of 3.20 {\AA}. There are 4 Mo and 16 P  atoms in the unit cell. All the P atoms can be divided into two types. The first type of P atoms is covalently bonded with three adjacent P atoms and forms a dipolar bond with one Mo atom. The other type of P atoms is covalently bonded with two neighboring P atoms and forms dipolar bonds with two Mo atoms. This bonding arrangement distorts the P atoms from their original in-plane positions of black phosphorus and imposes three-dimensionality in MoP$_4$. The crystal lattice only respects inversion $\mathcal{I}$, a two-fold rotation $\mathcal{\tilde{C}}_{2y} : (x,y,z) \Longrightarrow (-x, y, -z + \frac{1}{2})$, and a single glide mirror $\widetilde{\mathcal{M}}_y$ $: (x,y,z) \Longrightarrow (x, -y, z + \frac{1}{2})$ symmetries. Figure~\ref{Fig_CES}(g) shows the primitive cell Brillouin zone (BZ) and projected (001) surface BZ with marked $\widetilde{\mathcal{M}}_y$ mirror plane and the high-symmetry points.

\begin{figure}[ht]
\centering
\includegraphics[width=0.5\textwidth]{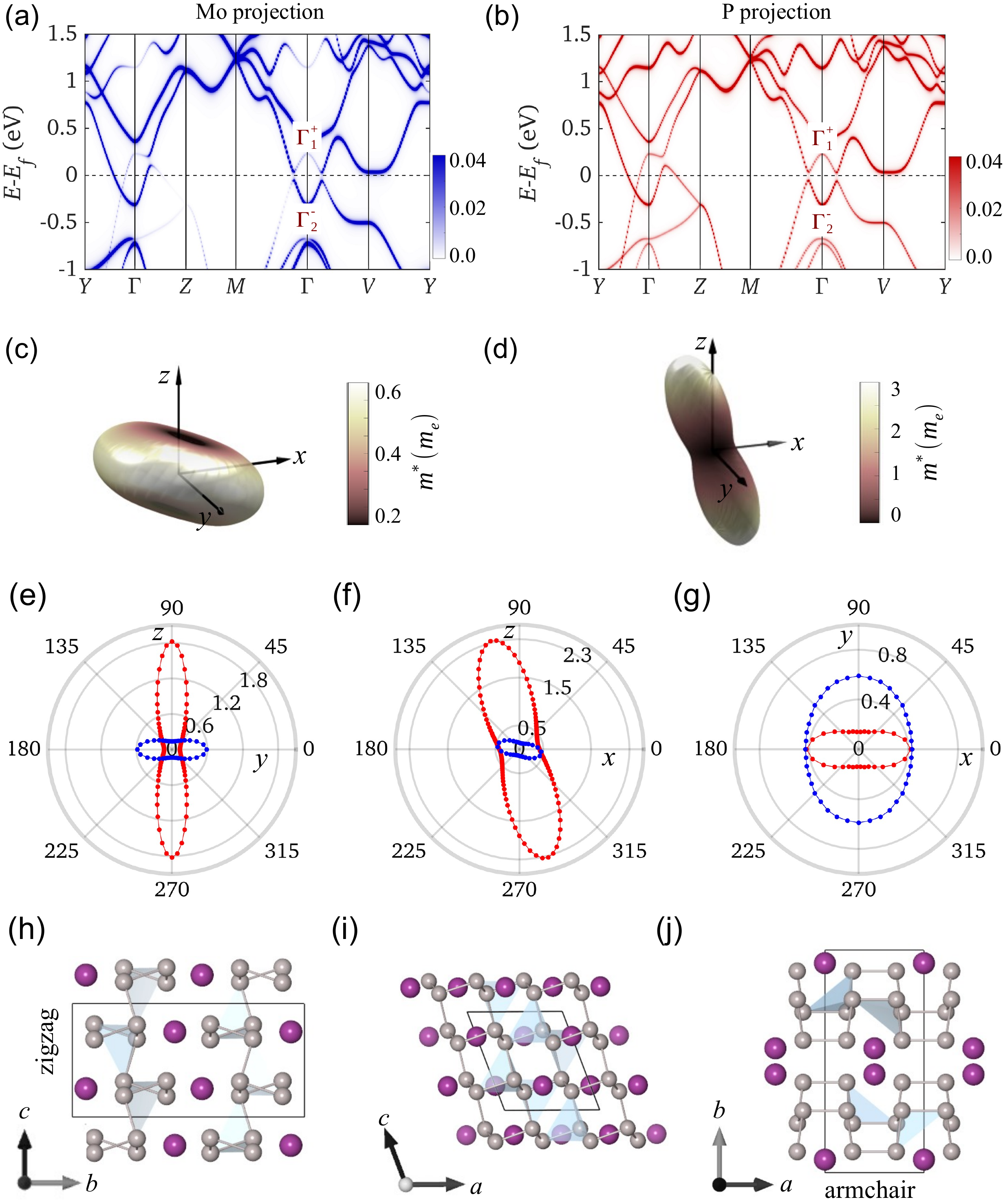}
\caption{{\bf Effective-mass anisotropy in MoP$_4$.} (a) Mo and (b) P atoms resolved band structure of MoP$_4$. The atomic weights are encoded in the color scale.
An electron-type Mo band ($\Gamma_2^-$) intersects the P resolved valence band ($\Gamma_1^+$) near the $\Gamma$ point. Calculated effective mass ellipsoid of (c) Mo and (d) P bands around the $\Gamma$ point. A highly anisotropic peanut-shaped variation of the effective mass revealed the anisotropic character of P bands. (e)-(g) Calculated effective masses at constant planes for the Mo band (blue color) and P band (red color). Angular direction is given from the horizontal axis in each panel. (h)-(j) Two-dimensional view of the MoP$_4$ crystal structure in the $b$-$c$, $a$-$c$, and $a$-$b$ planes. The zigzag and armchair directions are along the $c$ and $a$ directions, respectively. The P atomic layers are stacked along the $b$ direction. The effective mass of the P band is much larger in the zigzag direction than in the armchair direction and follows the characteristic anisotropic structure of phosphorene.} \label{Fig_Anisobands}
\end{figure}

\begin{table}[ht!]
\caption{Calculated effective masses of P ($\Gamma_1^+$) and Mo ($\Gamma_2^-$) bands in the $y-z$, $x-z$, and $x-y$ planes. Anisotropic ratio $\delta_{\{\Gamma_1^+,\Gamma_2^-\}} = \frac{m^*_{max}}{m^*_{min}}$, where $m^*_{max}$ and $m^*_{min}$ are maximum and minimum effective masses in each plane for P and Mo bands, is also given. The angular location of $m^*_{max}$ and $m^*_{min}$ added in parathesis is calculated w.r.t to the horizontal axis in each plane (see Fig.~\ref{Fig_Anisobands}). Effective masses are given in units of $m_e$.}
\resizebox{\linewidth}{!}{
\begin{tabular}{c c c c c c c}
\hline\hline
Plane & \multicolumn{3}{c}{P band ($\Gamma_1^+$) } & \multicolumn{3}{c}{Mo band ($\Gamma_2^-$)} \\
& $m^*_{max}$ & $m^*_{min}$ & $\delta_{\Gamma_1^+}$ & $m^*_{max}$ & $m^*_{min}$ & $\delta_{\Gamma_2^-}$ \\
\hline 
$y-z$ & $1.82~(90^\circ)$ & $0.13~(0^\circ)$ & $14$ & $0.59~(0^\circ)$ & $0.14~(90^\circ)$ & $4.21$ \\
$x-z$ & $2.34~(105^\circ)$ & $0.40~(15^\circ)$ & $5.85$ & $0.47~(165^\circ)$ & $0.13~(75^\circ)$ & $3.61$ \\
$x-y$ & $0.42~(0^\circ)$ & $0.14~(90^\circ)$ & $3$ & $0.59~(90^\circ)$ & $0.43~(0^\circ)$ & $1.37$ \\
\hline \hline
\end{tabular}
}
\label{tab1}
\end{table}

Figure~\ref{Fig_CES}(f) shows the calculated band structure of MoP$_4$ without SOC. It is semimetallic where the $\Gamma_1^+$ valence band intersects the $\Gamma_2^-$ conduction band with an inverted band ordering at the $\Gamma$ point. Particularly the valence and conduction band crossings stay ungapped along the $\Gamma-Y$ direction. There are additional two-fold band crossings along the $Z-M$ direction in the valence and conduction region of the band structure. Upon the inclusion of SOC, the band structure is locally gapped at each $k$ point, separating valence and conduction bands in the entire BZ (see the Supplemental Material~\cite{supplementary}). Nevertheless, the SOC-induced gap at the nodal points is less than 15 meV, which is slightly increased to a value of 25 meV when SOC artificially scaled to 500\%. This bandgap opening preserves the fermionology of the crossing bands so that SOC effects can be ignored. Moreover, the band crossings at $Z$ and $M$ points in the valence and conduction region stay robust, realizing nonsymmorphic symmetry-protected Dirac states in MoP$_4$~\cite{MoP4_HighPressure}.

To uncover lattice-driven anisotropic electronic properties and their connection to black phosphorus, we present the orbital-resolved band structure of MoP$_4$ in Figs.~\ref{Fig_Anisobands}(a)-(b). The conduction band is derived from the Mo atoms, whereas the valence band is composed of P atoms. These two bands of distinct atomic character cross in the vicinity of the Fermi level to generate a semimetallic state. The nature of the valence and conduction bands is further probed by calculating the orientation-dependent effective masses around the $\Gamma$ point (see Table~\ref{tab1}). In Figs.~\ref{Fig_Anisobands}(c) and \ref{Fig_Anisobands}(d), we present these results for the Mo and P bands, respectively, in the $x-y-z$ space. The effective mass associated with the Mo band is less anisotropic and forms a spheroid shape in space. Whereas the P band effective mass is highly anisotropic, forming a peanut-like shape and mimicking the anisotropic P lattice structure in MoP$_4$. The structure-to-mass-anisotropy relation is revealed in results shown on various plane cuts in Figs.~\ref{Fig_Anisobands}(e)-(g) and their associated crystal directions in Figs.~\ref{Fig_Anisobands}(h)-(j). Particularly, the P band effective mass (red markers) along the zigzag direction is higher than the armchair direction, similar to phosphorene bands. In contrast, the Mo bands (blue markers) exhibit maximum and minimum effective masses oriented perpendicular to the P bands. The Mo and P bands with oppositely oriented effective masses cross in the vicinity of the Fermi level to generate a semimetallic state.

\begin{figure}[t!]
\centering
\includegraphics[width=0.5\textwidth]{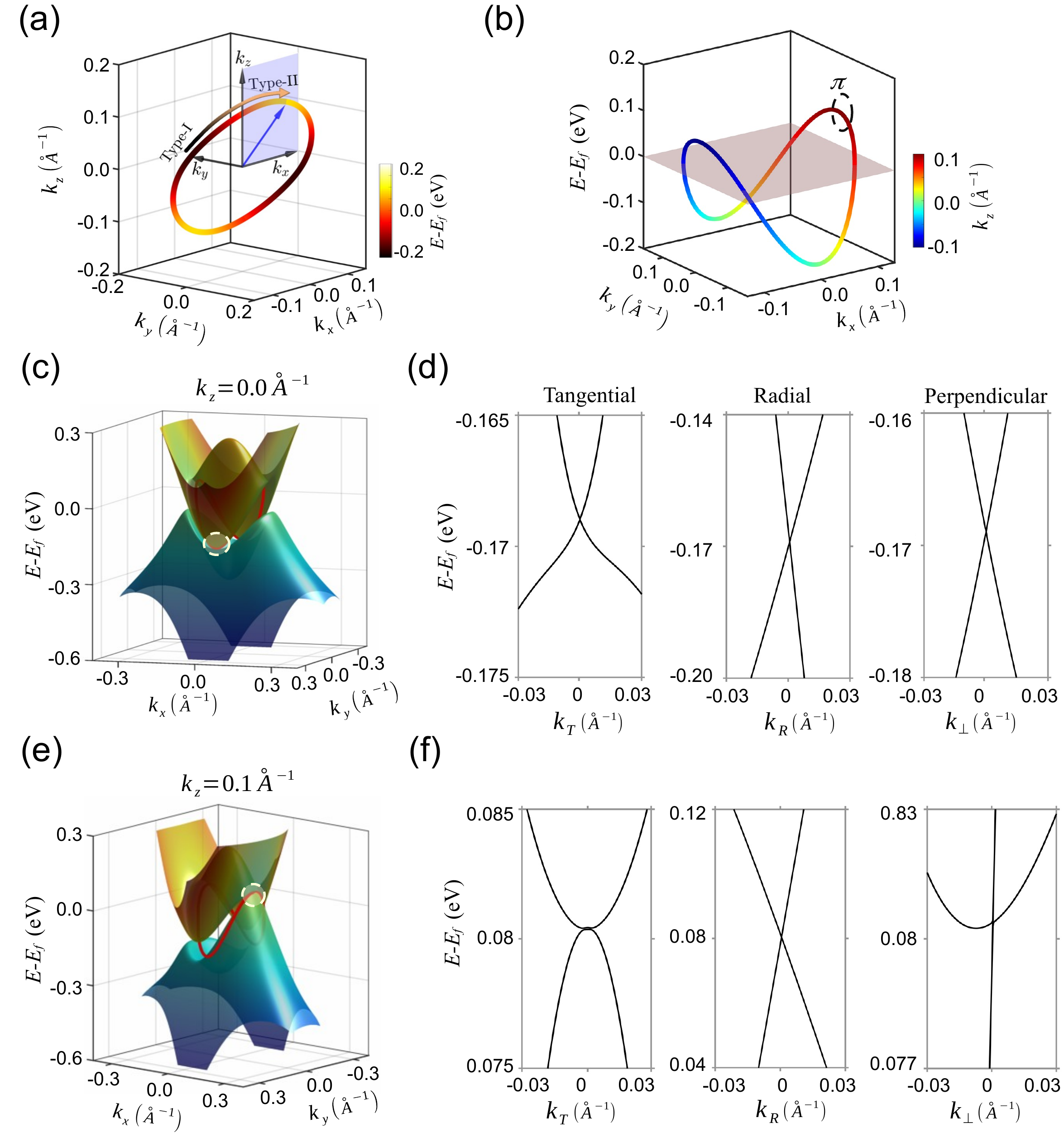}
\caption{{\bf Hybrid nodal line in MoP$_4$.} (a) The location of the nodal line in the reciprocal space. The color represents the nodal line energy. The arrow shows the transition from type-I to type-II points. The upper and lower portions of the nodal line across $\Gamma$ are related by inversion symmetry. (b) Energy-dependent nodal line configuration on $k_x-k_y$ plane. An extended energy region below and above the Fermi level (shown with the gray plane) is revealed. The color represents the $k_z$ location of the nodal line. (c),(e) Band structure in $k_x-k_y$ plane at (c) $k_z=0$ \r{A}$^{-1}$ and  (e) $k_z=0.1$ \r{A}$^{-1}$ with nodal line (red line). (d),(f) Energy dispersion along the tangential, radial, and perpendicular directions around the shaded points in (c),(e). At any point on the nodal line, $k_R$ defines radial direction, $k_T$ lies perpendicular to $k_R$ in the nodal line plane, and $k_\perp$ lies normal to the nodal line plane. From these energy dispersion cuts a type I character at $k_z=0$ \r{A}$^{-1}$ and a type II character (that is, along perpendicular directions both the crossings bands have the same sign of velocity) at $k_z=0.1$ \r{A}$^{-1}$ are revealed.}\label{Fig_topology}
\end{figure}

\subsection{Hybrid nodal line structure}

We now characterize the band crossings between Mo and P bands in Fig.~\ref{Fig_topology}. The gapless valence and conduction bands crossing points in $k_x-k_y-k_z$ space are presented in Fig.~\ref{Fig_topology}(a). These crossing points trace a line node that encloses the $\Gamma$ point. The line node is not hooked to the $k_y$ mirror plane but forms a $\mathcal{I}\mathcal{T}$ symmetric structure primarily located on the $k_x-k_z$ plane due to the presence of inversion symmetry. The symmetry protection of the nodal line is determined by calculating the Berry phase $\gamma=\oint dk \cdot A(k)$, where $A(k)=i\sum_n \langle u_{n,k}|\nabla u_{n,k}\rangle$ is the Berry connection of the occupied Bloch bands $|u_{n,k}\rangle$. For a generic closed $k$ loop encircling the line node, we obtain a Berry phase $\gamma=+\pi$. This results in a nontrivial winding number $\frac{\gamma}{\pi}=+1$, dictating the topological protection of the line node. The energy and momentum spread of the line node is shown in Figs.~\ref{Fig_topology}(a)-(b). The line node spans an energy range from $-0.17$ eV to $0.08$ eV that passes through the Fermi level. Such an extended energy range can enable spectroscopic verifications of the nodal line without fine-tuning the Fermi level in MoP$_4$.

Figures~\ref{Fig_topology}(c) and \ref{Fig_topology}(e) show the $E-k_x-k_y$ band dispersion at $k_z=0$ and $k_z=0.1$ \r{A}$^{-1}$ planes, respectively, with the line node shown in red color. The isolated nodal points (shaded white circle) lie on the $k_y$ axis at the $k_z=0$ plane, and on the $k_x$ axis at the $k_z=0.1$ plane. We further plot the energy dispersion away from these nodal points in Figs.~\ref{Fig_topology}(d) and \ref{Fig_topology}(f). Dispersing away from the nodal points, the two bands form nearly quadratic band crossing along the tangential direction and conical band crossings along the other two momentum directions on the $k_z=0$ plane. In contrast, the two bands have the same sign of velocity along one momentum direction (in this case, perpendicular direction to the nodal line plane) on the $k_z=0.1$ \r{A}$^{-1}$. These results clearly show that the line node is hybrid in character, having both type-I and type-II band crossings. Exploring the full momentum space location of type-I and type-II band crossings, we find that the type of band crossings is hooked to their lattice directions. In particular, the type-I nodal points lie along the armchair directions whereas the type-II band touchings lie along the zigzag direction of the P atoms. The nodal line dispersion continuously evolves from type I to type II in intermediate directions as shown with an arrow in Fig.~\ref{Fig_topology}(a) (see the Supplemental Material~\cite{supplementary} for details). Such a hybrid nodal structure arises due to crossings of anisotropic Mo and P bands with oppositely oriented effective masses, which generates a dispersive nodal line by displacing the crossing bands in opposite momentum directions. The large opposite displacement between the crossing bands drives type II dispersion along certain momentum directions (see Sec.~\ref{model}).

\begin{figure}[t!]
\centering
\includegraphics[width=0.48\textwidth]{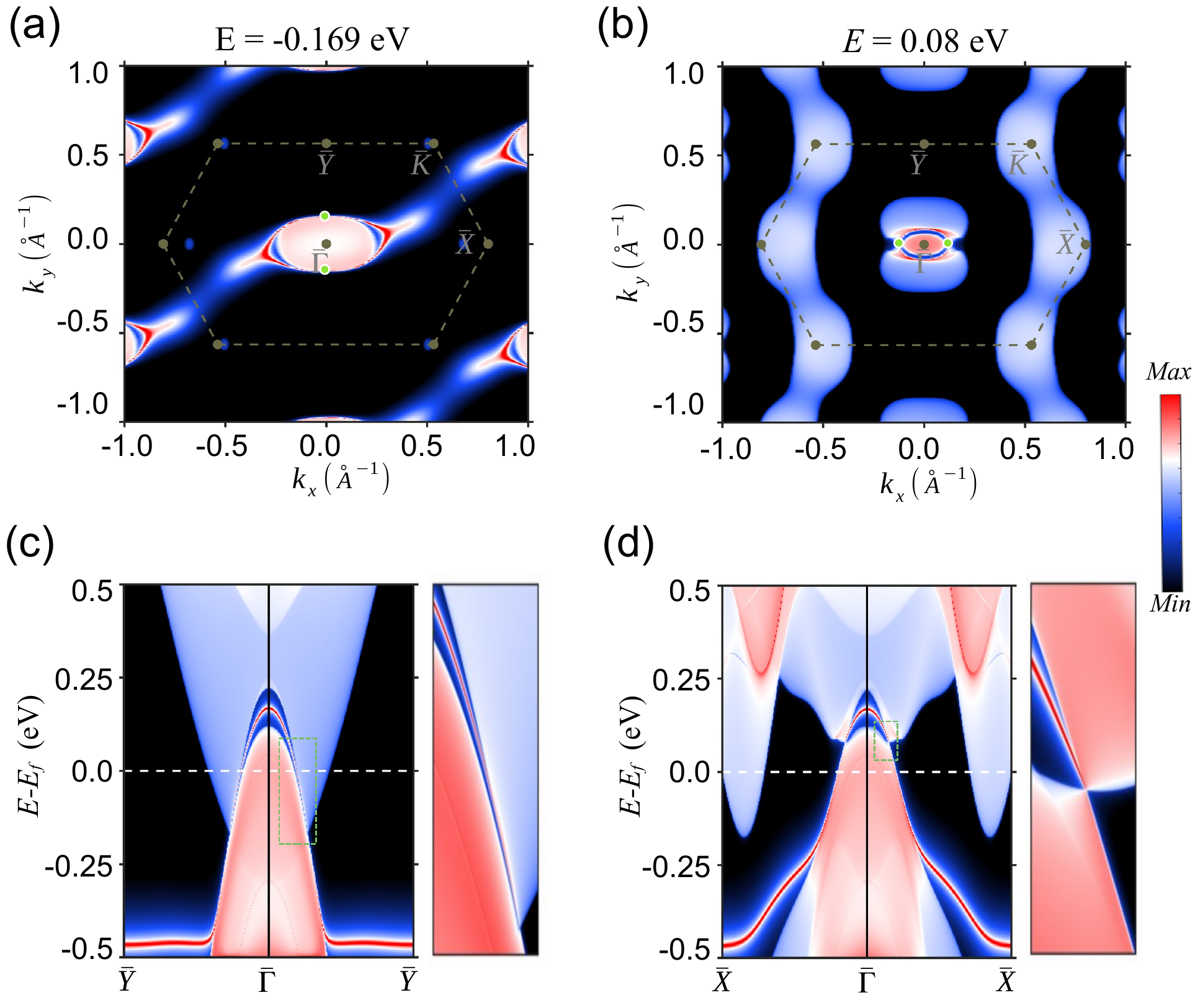}
\caption{{\bf Topological surface states in MoP$_4$.} Surface energy spectrum for the (001) surface of MoP$_4$ at constant energies (a) $E=-0.169$ eV and (b) $E=0.08$ eV obtained with a surface potential energy of -50 meV. The green dots in (a) and (b) mark the type-I and type-II projected band crossings along the high-symmetry axes, respectively. A fermi-arc-type connectively is revealed in the drumhead surface states at these iso-energy contours. Surface band structure of the (001) surface along the (c) $\bar{Y}$-$\bar{\Gamma}$-$\bar{Y}$ and (d) $\bar{X}$-$\bar{\Gamma}$-$\bar{X}$ directions. The left panel shows the closeup band structure highlighted in dashed boxes. The drumhead surface states lie inside the projected bulk band crossings.}\label{Fig_surfstate}
\end{figure}

\subsection{Topological surface states}
The existence of drumhead electron states inside or outside nodal-line projections on the crystal surface is the hallmark of topological nodal-line semimetals. To showcase these states, we present the calculated (001) surface band structure of MoP$_4$ in Fig.~\ref{Fig_surfstate}. In Figs.~\ref{Fig_surfstate}(a) and \ref{Fig_surfstate}(b), we illustrate the constant energy contours at $E=-0.169$ eV and $E=0.08$ eV as the representative case of type-I and type-II bulk nodal energies, respectively. These plots show a rich electronic structure with surface states located inside and outside the projected nodal line. The surface states connect two nodal points at a particular energy cut, forming a double Fermi-arc-type connectivity. These Fermi arc-type states are part of the highly dispersive drumhead surface states as seen in the surface band structure along $\overline{Y}-\overline{\Gamma}-\overline{Y}$ and $\overline{X}-\overline{\Gamma}-\overline{X}$ directions in Figs.~\ref{Fig_surfstate}(c) and \ref{Fig_surfstate}(d). Such drumhead electronic states and their Fermi-arc-type constant energy connectivity are unique to the dispersive hybrid line node and may serve as spectroscopic fingerprints for their experimental verifications.

\subsection{Effective Hamiltonian}\label{model} 

\begin{figure}[tbp]
\begin{center}
\includegraphics[width=0.48\textwidth]{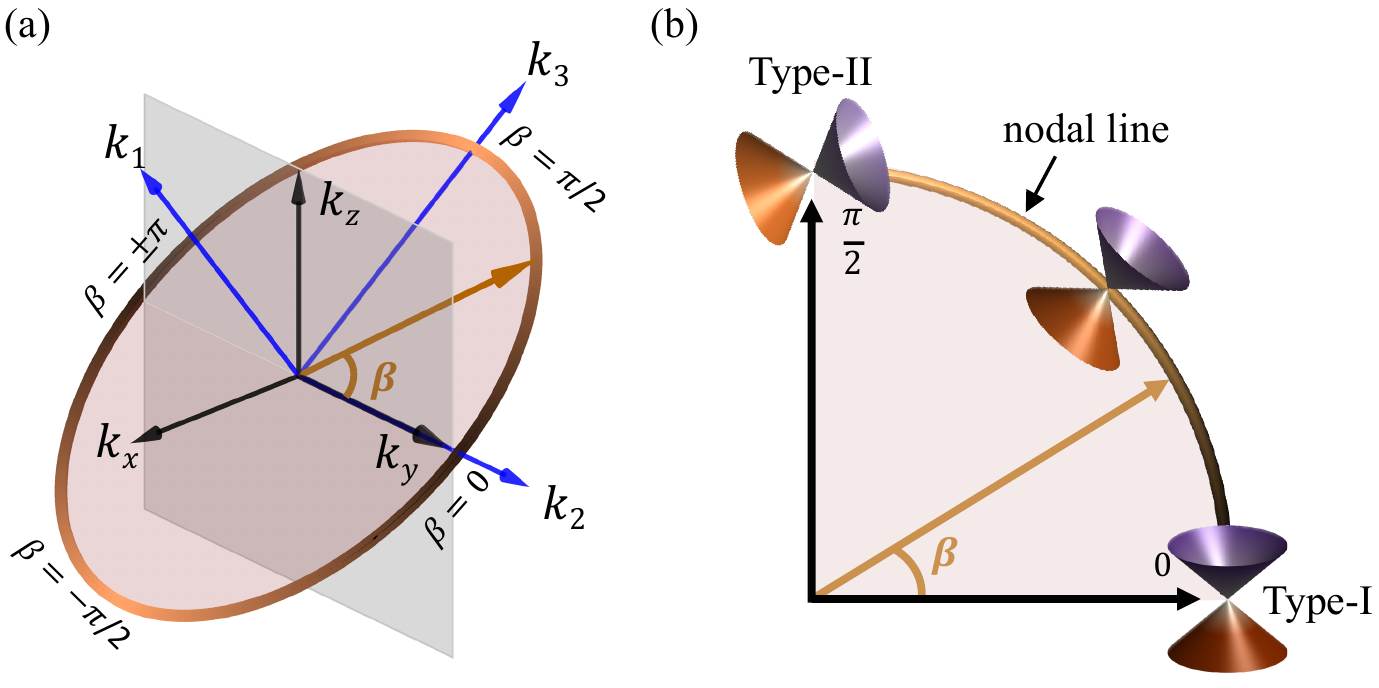}
\end{center}
\caption{{\bf Nodal ring configuration and energy dispersion.} (a) Nodal ring configuration in the global coordinates system ($k_x,k_y,k_z$).  
A new coordinate system ($k_1,k_2, k_3$) is defined where $k_2$ is along $k_y$, $k_1$ is perpendicular to the nodal-ring plane, and $k_3$ is perpendicular to $k_1$ \& $k_2$ and lie in the nodal-ring plane.
Nodal ring is parametrized by $\beta \in [-\pi,\pi]$. (b) Variation of nodal ring energy dispersion as a function of $\beta$ in the first quadrant. The energy dispersion evolves from type-I to type-II for $\beta=0$ to $\beta=\pi/2$.}
\label{basis}
\end{figure}

To better understand the underlying mechanism of mass-anisotropy driven hybrid nodal line, we drive the low-energy effective Hamiltonian using the theory of invariants~\cite{singh2018saddle,Model_TI}. The first-principles results indicate that anisotropic $\Gamma^{-}_2$ and $\Gamma^{+}_1$ bands cross to generate the hybrid nodal line. A $k\cdot p$ Hamiltonian around $\Gamma$ is constrained by time-reversal symmetry $\Theta: H^*(\vb{k})=H(-\vb{k})$, glide mirror $\widetilde{\mathcal{M}}_y: H (\vb{k}) = H(M_y\vb{k})$ and inversion symmetry $\mathcal{I}: \sigma_z H(\vb{k})\sigma_z = H(-\vb{k}$),  where $\sigma_z$ is a Pauli matrix. 
Based on these symmetry constraints, the spinless two-band $k\cdot p$  Hamiltonian takes the form
\begin{equation}\label{Eq_Ham}
H(\vb{k}) = \mqty( \varepsilon^{(\text{u})}(\vb{k}) & \Delta(\vb{k}) \\ \Delta^*(\vb{k}) & \varepsilon^{(\text{l})}(\vb{k}) ),
\end{equation} 
where $\varepsilon^{(\text{u})}$ and $\varepsilon^{(\text{l})}$ are associated with $\Gamma^{-}_2$ and $\Gamma^{+}_1$ bands with 
\begin{align}
\varepsilon^{(\text{u,l})}(\vb{k}) &= \pm \left\{ \sum_{i=x,y,z} \left(\frac{1}{2} \alpha^{(\text{u,l})}_{i} k_i^2 \right) + \alpha^{(\text{u,l})}_{xz} k_x k_z + \frac{1}{2} \epsilon_0 \right\},  \\
\Delta(\vb{k}) &= -i V \left( v_x k_x - v_z k_z \right). 
\end{align} 
Here all the parameters are considered real.  To have concave upward and downward curves for $\varepsilon^{(\text{u})}$ and $\varepsilon^{(\text{l})}$, the inverse of absolute effective masses $\alpha^{(\text{u})}_{i},\alpha^{(\text{l})}_{i}>0$ for $i=x,y,z$ (here $\hbar =1$). The $\alpha^{(\text{u,l})}_{xz}$ term is introduced to break the mirror symmetry on the $k_x$ and $k_z$ planes, giving tilted effective mass eclipses as found in our calculations. $\epsilon_0< 0$ gives the band inversion between the two bands at $\Gamma$. 
%
%
$\Delta(\vb{k})$ describes the interband coupling with strength $\abs{V\vb{k}}$. The dimensionless parameters $v_x$ and $v_z$ with a constraint $v_x^2+v_z^2=1$ describe the anisotropy due to interband coupling.  
The eigen energies associated with the Hamiltonian (Eq.~\ref{Eq_Ham}) are
\begin{equation}\label{eigenval}
    E_{\pm}( \vb{k}) = \xi^{(+)}(\vb{k}) \pm \sqrt{\xi^{(-)}(\vb{k})^2 + \abs{\Delta(\vb{k})}^2},
\end{equation}
where $\xi^{(\pm)}(\vb{k}) = \frac{1}{2} \left[ \varepsilon^{(\text{u})}(\vb{k}) \pm \varepsilon^{(\text{l})}(\vb{k})\right]$. For later use, we define 
$\alpha_{X}^{(\text{u,l})} = v_z^2 \alpha_x^{(\text{u,l})} + v_x^2 \alpha_z^{(\text{u,l})} + 2 v_x v_z  \alpha_{xz}^{(\text{u,l})}$, $\alpha_{Z}^{(\text{u,l})} =v_x v_z (\alpha_x^{(\text{u,l})} - \alpha_z^{(\text{u,l})})+ (v_x^2 -v_z^2) \alpha_{xz}^{(\text{u,l})}$. 
Since the two crossing bands have equal mirror eigenvalues, their nodal crossings are not protected by $\widetilde{\mathcal{M}}_y$. Instead, the inversion symmetry $\mathcal{I}$ can protect a line nodal which is determined by the conditions $\xi^{(-)}(\vb{k}) =0$ and $\Delta(\vb{k})=0$. The nodal ring for $\beta \in (-\pi,\pi]$ [see Fig.~\ref{basis}(b)] is positioned at
\begin{equation}
\vb{k}_{\textrm{node}}(\beta) = \left( v_z K_{X} \sin(\beta), \pm K_y \cos(\beta), v_x K_{X} \sin(\beta) \right) ,
\end{equation} 
where $K_{X} =\sqrt{\frac{\abs{\varepsilon_0}}{\alpha_X}}$ and $K_{y} =\sqrt{\frac{\abs{\varepsilon_0}}{\alpha_y}}$. 

To describe the local energy dispersion associated with nodal crossings for different $\beta$, we define a new coordinate system $(k_1,k_2,k_3)$ as shown in Fig.~\ref{basis}(a). The transformation of  $(k_1,k_2,k_3) \rightarrow (k_x,k_y,k_z)$ is $k_x = v_x k_1 + v_z k_3$, $k_y=k_2$, and $k_z = -v_z k_1 + v_x k_3$. Here $k_1$ is perpendicular to the nodal ring plane ($\Delta(\vb{k})=0$). Taking $k_2 = K_y \cos \beta$ and $k_3 = K_X \sin \beta $ on the nodal ring, the dispersions of $\varepsilon^{(\text{u})}$ and $\varepsilon^{(\text{l})}$ along $k_1$ takes the form 
\begin{equation}
    \varepsilon^{(\text{u,l})}(k_1) =  \pm  \left\{ \frac{1}{2} A^{(\text{u,l})} k_1^2 +  \alpha_{Z}^{(\text{u,l})} K_X  \sin(\beta) k_1  \right\} + E_{\textrm{node}}(\beta).
    \label{dispersion}
\end{equation}
where $A^{(\text{u,l})}= \alpha_x^{(\text{u,l})} + \alpha_z^{(\text{u,l})} - \alpha_X^{(\text{u,l})}$. $E_{\textrm{node}}(\beta) = \frac{1}{2}  K_{X}^2 \delta \alpha_{X} \sin^2(\beta) + \frac{1}{2} K_y^2 \delta \alpha_y \cos^2(\beta) $ is the energy of the nodal ring, which agrees with our first-principles results (Fig.~\ref{Fig_topology}(b)) for $K_X^2 \delta \alpha_X > K_y^2 \delta \alpha_y $. The energy dispersions $\varepsilon^{(\text{u,l})}(k_1)$ shows shifted parabolic bands that cross at $k_1=0$ with shifts proportional to $\alpha_z^{(\text{u,l})}\sin(\beta)$. Notably, the band shifts increase with $\beta$. When the band shifts are opposite, $\alpha_{Z}^{(\text{u})} \alpha_{Z}^{(\text{l})}<0$, a type II feature emerges (Fig.~\ref{basis}(b)). Below we show that this is the necessary condition for realizing a type-II band crossing.

Including the interband coupling, $\Delta(\vb{k})= -i V k_1$, the slopes of $E_{\pm}( \vb{k})$ along $k_1$ are 
\begin{equation}
   \pdv{k_1} E_{\pm}(\vb{k}) = K_{X} \delta \alpha_{Z} \sin(\beta) \pm \sqrt{\left[ K_{X}\alpha_{Z}  \sin(\beta) \right]^2 + V^2},
\end{equation}
where $ \alpha_{Z} = \frac{1}{2} \left( \alpha_z^{(\text{u})} + \alpha_z^{(\text{l})} \right)$ and $ \delta \alpha_{Z} = \frac{1}{2}\left( \alpha_z^{(\text{u})} - \alpha_z^{(\text{l})} \right)$.
The condition for a type II nodal dispersion is 
\begin{equation}
   \abs{ \delta \alpha_{Z} K_{X} \sin(\beta)} > \sqrt{\left[\alpha_{Z}  K_{X} \sin(\beta) \right]^2 + V^2},
\end{equation}
which reduces to 
\begin{equation}
   \left(- \alpha_{Z}^{(\text{u})} \alpha_{Z}^{(\text{l})} \right) \sin^2(\beta) > \left(\frac{V}{K_{X}} \right)^2.
       \label{maineq}
\end{equation}

Notably, $\alpha_{Z}^{(\text{u,l})} $ characterizes the dressed mass anisotropy in the $k_x$-$k_z$ plane for the two bands. The necessary condition for realizing a type II band dispersion is that the mass anisotropies of the crossing bands should be opposite, $\alpha_{Z}^{(\text{u})} \alpha_{Z}^{(\text{l})}<0$. For small $\beta$ and finite $V$, Eq. (\ref{maineq}) can not be satisfied. This can be understood as for $\beta=0$, there is no $k$-shift in Eq.~(\ref{dispersion}), and the band crossing should be type-I. However, when $\beta$ approaches $\pm \pi/2$, a type-II band crossing can be realized (Fig.~\ref{basis}(b)) in the $k_1$ direction provided bands should have large mass anisotropy : $\abs{\beta}> \arcsin\left(\abs{V}/K_X \sqrt{ -\alpha_{Z}^{(\text{u})} \alpha_{Z}^{(\text{l})}} \right)$. The dispersion along $k_2$ always remains type-I as found in our first-principles results. 

We emphasize that Eq. (\ref{maineq}) is the main result of our work. Taking $\sin^2(\beta)=1$ describes the condition for a hybrid nodal ring. A large opposite mass anisotropy of the crossing bands is an ingredient to realizing a hybrid nodal line in materials. 
More precisely, the anisotropy includes the effect of particle-hole asymmetry between the bands. When the conduction and the valence bands have opposite mass anisotropies as found in $M$P$_4$, the crossing bands form a hybrid nodal line. 

\subsection{Materials tunability and phase transition}

\begin{figure}[t!]
\centering
\includegraphics[width=0.5\textwidth]{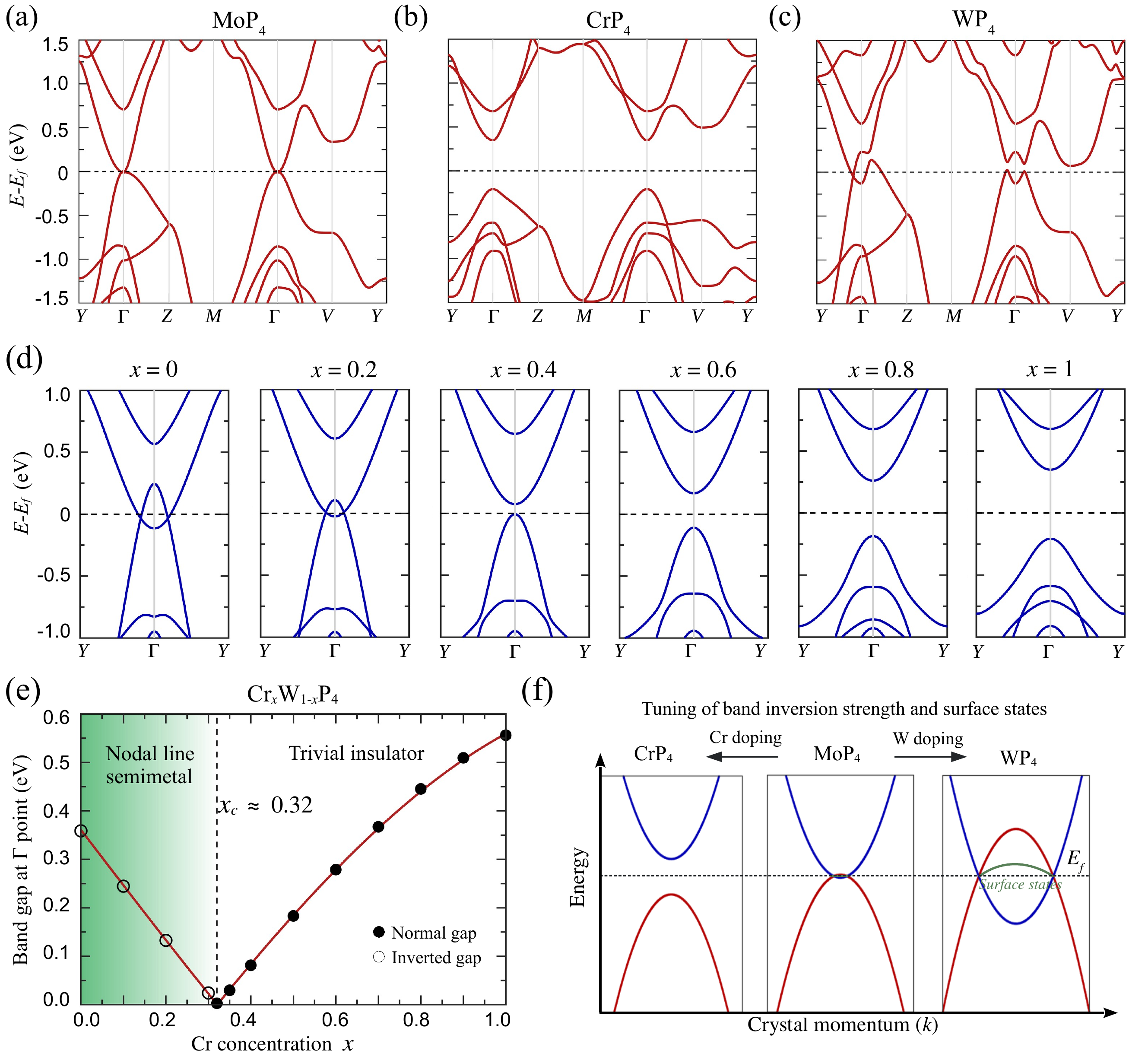}
\caption{{\bf Hybrid nodal-line to insulator transition and phase tunability.} Band structures of (a) MoP$_4$, (b) CrP$_4$, and (c) WP$_4$ obtained using the HSE hybrid exchange-correlation functional.  (d) Calculated band structure of Cr$_x$W$_{1-x}$P$_4$ along $Y-\Gamma-Y$ direction for different $x$ values. (e) Variation of energy gap at $\Gamma$ for Cr$_x$W$_{1-x}$P$_4$ as a function of $x$. The related topological state is highlighted. The red line is a guide to the eye.  (f) Schematic representation of calculated hybrid-functional band structure of representative compounds in the $M$P$_4$ family. Band inversion strength increases in WP$_4$ whereas CrP$_4$ realizes a fully gapped state.} \label{Fig_phasetuning}
\end{figure}

We now consider the robustness of the hybrid node line and demonstrate a semimetal-to-insulator transition in $M$P$_4$. Owing to the presence of $M$$-d$ bands, the electronic correlations may play an important role in dictating the nature of the nodal line in $M$P$_4$ since they could shape the energy dispersion and effective masses of the crossings bands. In Fig.~\ref{Fig_phasetuning}(a), we present the band structure of MoP$_4$ obtained with HSE06 hybrid functional that incorporates a part of the exact Fock exchange~\cite{heyd2006hybrid}. The calculated energy dispersions of valence and conduction bands stay preserved although their band overlap is reduced compared to GGA results. On exploring the mass anisotropies and nodal line energy dispersion, we find that the hybrid nodal line is quite robust to the changes to the exchange-correlation functionals (see the Supplemental Material~\cite{supplementary}). However, due to reduced overlap between the valence and conduction bands, MoP$_4$ lies close to a semimetal-to-insulator transition point.

In order to discuss the semimetal-to-insulator transition and the tunability of hybrid nodal line, it is useful to calculate the HSE06 band structures of CrP$_4$ and WP$_4$ (Figs.~\ref{Fig_phasetuning}(b) and \ref{Fig_phasetuning}(c)). 
Since CrP$_4$ and WP$_4$ are isostructural to MoP$_4$, they show similar mass anisotropies in the valence and conduction bands as seen in MoP$_4$~\cite{TmP4_CS,MoP4_HighPressure,TmP4_CS_WP4}. However, the magnitude of effective masses and anisotropy are materials dependent owing to distinct interband coupling effects (see the Supplemental Material~\cite{supplementary} for details). 
This can be further understood based on the different structural parameters of $M$P$_4$. Due to the different effective radii of the $M$ atoms, the lattice constant is decreased in CrP$_4$ ($a=5.19$ \r{AA}, $b=10.16$ \r{AA}, and $c=5.77$ \r{AA}) and increased in WP$_4$ ($a=5.34$ \r{AA}, $b=11.19$ \r{AA}, and $c=5.87$ \r{AA}) as compared to the MoP$_4$. The change in lattice parameters results in different crystal-field effects and electronic states in CrP$_4$ and WP$_4$. Particularly, CrP$_4$ realizes an insulator state with a band gap of 0.54 eV at the $\Gamma$ point, whereas WP$_4$ forms a hybrid nodal line with an increased band inversion strength. These results demonstrate that CrP$_4$ and WP$_4$ lie electronically on the opposite sides of MoP$_4$. Alternatively, the Cr doping in MoP$_4$ will be expected to reduce the valence and conduction bands overlap and push the material toward the trivial insulator state. In contrast, W doping in MoP$_4$ will increase the band overlap and realize a hybrid nodal line with increased band inversion strength.

Figure~\ref{Fig_phasetuning}(d) shows energy dispersion of Cr$_x$W$_{1-x}$P$_4$ alloys for various Cr concentrations $x$ calculated along the $k_y$ axis. The virtual crystal approximation is adopted to model the band structure using the Wannier Hamiltonian obtained with the HSE06 functional. When $x=0$, a clear overlap between $M$$-d$ and P$-p$ states is seen so that the system realizes a nodal line semimetal state. As $x$ is increased, the $M$-$d$ and P-$p$ state move in opposite energy directions and touch at the $\Gamma$ point for $x_c=0.32$. With a further increase in $x$, the overlap between these states vanishes, and the system transitions to an insulator state. 
Notably, we calculate the orientation-dependent effective masses of valence and conduction bands and nodal line dispersion of Cr$_x$W$_{1-x}$P$_4$ for $x=0.2$. The mass anisotropy of bands and the hybrid nature of the nodal line dispersion remain preserved, showing that these features are robust in $M$P$_4$ materials. 
The evolution of bandgap of Cr$_x$W$_{1-x}$P$_4$ as a function of $x$ at $\Gamma$ is shown in Fig~\ref{Fig_phasetuning}(e). A hybrid semimetal-to-insulator transition can thus be achieved by varying the Cr concentration. A similar tunable system can be achieved by doping MoP$_4$ with Cr atoms (see Fig.~\ref{Fig_phasetuning}(f)). Since Cr, W, and Mo atoms are isovalent and form isostructural materials with similar mass anisotropies, one can create Cr$_x$(Mo, W)$_{1-x}$P$_4$ material that can be tuned among various topological states.

\section{Discussion}

Understanding the structure-to-property relationship lies at the heart of physics research and provides concepts for engineering new device design~\cite{SPR_2DXinming2017}. It remains largely unexplored in topological materials, even though the numbers and energy dispersions of nontrivial states depend on the structural and positional motifs of materials. Based on the first-principles calculations and $k \cdot p$ theory, we have systematically demonstrated how the lattice geometry-driven effective mass anisotropies result in unique hybrid nodal line states in black phosphorus materials $M$P$_4$. Taking MoP$_4$ as an explicit example, we show that it retains characteristic anisotropic energy dispersion of the phosphorene and harbors a single hybrid nodal line between Mo-$d$ and P-$p$ states. The nodal line constitutes both the type-I and type-II band crossings and spreads across the Fermi level around the $\Gamma$ point. We reveal that the P bands have a peanut-type effective mass variation such that the large and small effective masses orient along the zigzag and armchair directions of the P atoms, respectively. The anisotropic P band crosses with the Mo-$d$ band to form the hybrid nodal line where the type-I and type-II nodal points are hooked to the armchair and zigzag lattice directions. Through in-depth symmetry and Hamiltonian analysis, we show that the strong mass anisotropy in the crossing bands is a necessary condition to form the hybrid nodal line in $M$P$_4$. We further propose the realization of the hybrid nodal line state with increased inversion strength and a metal-to-insulator transition through isoelectronic chemical substitutions such as Cr$_x$(W, Mo)$_{1-x}$P$_4$. $M$P$_4$ materials have been synthesized in experiments and explored in connection with high diffusion anisotropy of sodium or other metal ions in the phosphorous layers~~\cite{TMBP_Gong2018,BPNa_Sun2015,cheng2017sodium,TmP4_CS,MoP4_HighPressure,TmP4_CS_WP4,TmP4_Khan2020}. More recent work discusses the high-pressure synthesis of MoP$_4$ and reveals its semimetallic nature with high-mobility electrons~\cite{MoP4_HighPressure}. Since the single crystalline samples are grown, and the hybrid nodal line runs across the Fermi level, the experimental validation of the nodal line could be done, for example, in photoemission experiments without any Fermi level tuning. Moreover, orientation-dependent effective masses can be determined from the transport experiments. Their anisotropic behavior could serve as experimental fingerprints to determine the nodal line dispersion in topological semimetals. These facts highlight that $M$P$_4$ materials constitute an ideal platform to explore mass-anisotropy-driven hybrid nodal states and advance our understanding of the role of lattice geometry in shaping the topological state dispersion in materials.

\section*{Acknowledgements} 
This work is supported by the Department of Atomic Energy of the Government of India under Project No. 12-R$\&$D-TFR-5.10-0100 and benefited from the computational resources of TIFR Mumbai. S.M.H. is supported by the NSTC-AFOSR Taiwan program on Topological and Nanostructured Materials, Grant No. 110-2124-M-110-002-MY3.

\bibliography{MoP4}

\begin{thebibliography}{57}%
\makeatletter
\providecommand \@ifxundefined [1]{%
 \@ifx{#1\undefined}
}%
\providecommand \@ifnum [1]{%
 \ifnum #1\expandafter \@firstoftwo
 \else \expandafter \@secondoftwo
 \fi
}%
\providecommand \@ifx [1]{%
 \ifx #1\expandafter \@firstoftwo
 \else \expandafter \@secondoftwo
 \fi
}%
\providecommand \natexlab [1]{#1}%
\providecommand \enquote  [1]{``#1''}%
\providecommand \bibnamefont  [1]{#1}%
\providecommand \bibfnamefont [1]{#1}%
\providecommand \citenamefont [1]{#1}%
\providecommand \href@noop [0]{\@secondoftwo}%
\providecommand \href [0]{\begingroup \@sanitize@url \@href}%
\providecommand \@href[1]{\@@startlink{#1}\@@href}%
\providecommand \@@href[1]{\endgroup#1\@@endlink}%
\providecommand \@sanitize@url [0]{\catcode `\\12\catcode `\$12\catcode
  `\&12\catcode `\#12\catcode `\^12\catcode `\_12\catcode `\%12\relax}%
\providecommand \@@startlink[1]{}%
\providecommand \@@endlink[0]{}%
\providecommand \url  [0]{\begingroup\@sanitize@url \@url }%
\providecommand \@url [1]{\endgroup\@href {#1}{\urlprefix }}%
\providecommand \urlprefix  [0]{URL }%
\providecommand \Eprint [0]{\href }%
\providecommand \doibase [0]{https://doi.org/}%
\providecommand \selectlanguage [0]{\@gobble}%
\providecommand \bibinfo  [0]{\@secondoftwo}%
\providecommand \bibfield  [0]{\@secondoftwo}%
\providecommand \translation [1]{[#1]}%
\providecommand \BibitemOpen [0]{}%
\providecommand \bibitemStop [0]{}%
\providecommand \bibitemNoStop [0]{.\EOS\space}%
\providecommand \EOS [0]{\spacefactor3000\relax}%
\providecommand \BibitemShut  [1]{\csname bibitem#1\endcsname}%
\let\auto@bib@innerbib\@empty
\bibitem [{\citenamefont {Hasan}\ and\ \citenamefont
  {Kane}(2010)}]{hasan2010colloquium}%
  \BibitemOpen
  \bibfield  {author} {\bibinfo {author} {\bibfnamefont {M.~Z.}\ \bibnamefont
  {Hasan}}\ and\ \bibinfo {author} {\bibfnamefont {C.~L.}\ \bibnamefont
  {Kane}},\ }\bibfield  {title} {\bibinfo {title} {Colloquium: Topological
  insulators},\ }\href {https://doi.org/10.1103/RevModPhys.82.3045} {\bibfield
  {journal} {\bibinfo  {journal} {Rev. Mod. Phys.}\ }\textbf {\bibinfo {volume}
  {82}},\ \bibinfo {pages} {3045} (\bibinfo {year} {2010})}\BibitemShut
  {NoStop}%
\bibitem [{\citenamefont {Bansil}\ \emph {et~al.}(2016)\citenamefont {Bansil},
  \citenamefont {Lin},\ and\ \citenamefont {Das}}]{bansil2016colloquium}%
  \BibitemOpen
  \bibfield  {author} {\bibinfo {author} {\bibfnamefont {A.}~\bibnamefont
  {Bansil}}, \bibinfo {author} {\bibfnamefont {H.}~\bibnamefont {Lin}},\ and\
  \bibinfo {author} {\bibfnamefont {T.}~\bibnamefont {Das}},\ }\bibfield
  {title} {\bibinfo {title} {Colloquium: Topological band theory},\ }\href
  {https://doi.org/10.1103/RevModPhys.88.021004} {\bibfield  {journal}
  {\bibinfo  {journal} {Rev. Mod. Phys.}\ }\textbf {\bibinfo {volume} {88}},\
  \bibinfo {pages} {021004} (\bibinfo {year} {2016})}\BibitemShut {NoStop}%
\bibitem [{\citenamefont {Armitage}\ \emph {et~al.}(2018)\citenamefont
  {Armitage}, \citenamefont {Mele},\ and\ \citenamefont
  {Vishwanath}}]{RMP_WeylAshvin}%
  \BibitemOpen
  \bibfield  {author} {\bibinfo {author} {\bibfnamefont {N.~P.}\ \bibnamefont
  {Armitage}}, \bibinfo {author} {\bibfnamefont {E.~J.}\ \bibnamefont {Mele}},\
  and\ \bibinfo {author} {\bibfnamefont {A.}~\bibnamefont {Vishwanath}},\
  }\bibfield  {title} {\bibinfo {title} {Weyl and dirac semimetals in
  three-dimensional solids},\ }\href
  {https://doi.org/10.1103/RevModPhys.90.015001} {\bibfield  {journal}
  {\bibinfo  {journal} {Rev. Mod. Phys.}\ }\textbf {\bibinfo {volume} {90}},\
  \bibinfo {pages} {015001} (\bibinfo {year} {2018})}\BibitemShut {NoStop}%
\bibitem [{\citenamefont {Singh}\ \emph {et~al.}(2022)\citenamefont {Singh}, ,
  \citenamefont {Lin},\ and\ \citenamefont {Bansil}}]{singh2022topology}%
  \BibitemOpen
  \bibfield  {author} {\bibinfo {author} {\bibfnamefont {B.}~\bibnamefont
  {Singh}}, , \bibinfo {author} {\bibfnamefont {H.}~\bibnamefont {Lin}},\ and\
  \bibinfo {author} {\bibfnamefont {A.}~\bibnamefont {Bansil}},\ }\bibfield
  {title} {\bibinfo {title} {Topology and symmetry in quantum materials},\
  }\href {https://doi.org/https://doi.org/10.1002/adma.202201058} {\bibfield
  {journal} {\bibinfo  {journal} {Adv. Mater.}\ ,\ \bibinfo {pages} {2201058}}
  (\bibinfo {year} {2022})}\BibitemShut {NoStop}%
\bibitem [{\citenamefont {Fu}(2011)}]{fu2011topological}%
  \BibitemOpen
  \bibfield  {author} {\bibinfo {author} {\bibfnamefont {L.}~\bibnamefont
  {Fu}},\ }\bibfield  {title} {\bibinfo {title} {Topological crystalline
  insulators},\ }\href {https://doi.org/10.1103/PhysRevLett.106.106802}
  {\bibfield  {journal} {\bibinfo  {journal} {Phys. Rev. Lett.}\ }\textbf
  {\bibinfo {volume} {106}},\ \bibinfo {pages} {106802} (\bibinfo {year}
  {2011})}\BibitemShut {NoStop}%
\bibitem [{\citenamefont {Hsieh}\ \emph {et~al.}(2012)\citenamefont {Hsieh},
  \citenamefont {Lin}, \citenamefont {Liu}, \citenamefont {Duan}, \citenamefont
  {Bansil},\ and\ \citenamefont {Fu}}]{hsieh2012topological}%
  \BibitemOpen
  \bibfield  {author} {\bibinfo {author} {\bibfnamefont {T.~H.}\ \bibnamefont
  {Hsieh}}, \bibinfo {author} {\bibfnamefont {H.}~\bibnamefont {Lin}}, \bibinfo
  {author} {\bibfnamefont {J.}~\bibnamefont {Liu}}, \bibinfo {author}
  {\bibfnamefont {W.}~\bibnamefont {Duan}}, \bibinfo {author} {\bibfnamefont
  {A.}~\bibnamefont {Bansil}},\ and\ \bibinfo {author} {\bibfnamefont
  {L.}~\bibnamefont {Fu}},\ }\bibfield  {title} {\bibinfo {title} {Topological
  crystalline insulators in the snte material class},\ }\href
  {https://doi.org/10.1038/ncomms1969} {\bibfield  {journal} {\bibinfo
  {journal} {Nat. Commun.}\ }\textbf {\bibinfo {volume} {3}},\ \bibinfo {pages}
  {982} (\bibinfo {year} {2012})}\BibitemShut {NoStop}%
\bibitem [{\citenamefont {Fang}\ and\ \citenamefont {Fu}(2019)}]{RTCI_Fu2019}%
  \BibitemOpen
  \bibfield  {author} {\bibinfo {author} {\bibfnamefont {C.}~\bibnamefont
  {Fang}}\ and\ \bibinfo {author} {\bibfnamefont {L.}~\bibnamefont {Fu}},\
  }\bibfield  {title} {\bibinfo {title} {New classes of topological crystalline
  insulators having surface rotation anomaly},\ }\href
  {https://doi.org/10.1126/sciadv.aat2374} {\bibfield  {journal} {\bibinfo
  {journal} {Sci. Adv.}\ }\textbf {\bibinfo {volume} {5}},\ \bibinfo {pages}
  {eaat2374} (\bibinfo {year} {2019})}\BibitemShut {NoStop}%
\bibitem [{\citenamefont {Bradlyn}\ \emph {et~al.}(2016)\citenamefont
  {Bradlyn}, \citenamefont {Cano}, \citenamefont {Wang}, \citenamefont
  {Vergniory}, \citenamefont {Felser}, \citenamefont {Cava},\ and\
  \citenamefont {Bernevig}}]{Multi_Fermion}%
  \BibitemOpen
  \bibfield  {author} {\bibinfo {author} {\bibfnamefont {B.}~\bibnamefont
  {Bradlyn}}, \bibinfo {author} {\bibfnamefont {J.}~\bibnamefont {Cano}},
  \bibinfo {author} {\bibfnamefont {Z.}~\bibnamefont {Wang}}, \bibinfo {author}
  {\bibfnamefont {M.~G.}\ \bibnamefont {Vergniory}}, \bibinfo {author}
  {\bibfnamefont {C.}~\bibnamefont {Felser}}, \bibinfo {author} {\bibfnamefont
  {R.~J.}\ \bibnamefont {Cava}},\ and\ \bibinfo {author} {\bibfnamefont
  {B.~A.}\ \bibnamefont {Bernevig}},\ }\bibfield  {title} {\bibinfo {title}
  {Beyond dirac and weyl fermions: Unconventional quasiparticles in
  conventional crystals},\ }\href {https://doi.org/10.1126/science.aaf5037}
  {\bibfield  {journal} {\bibinfo  {journal} {Science}\ }\textbf {\bibinfo
  {volume} {353}},\ \bibinfo {pages} {aaf5037} (\bibinfo {year}
  {2016})}\BibitemShut {NoStop}%
\bibitem [{\citenamefont {Xiao}\ \emph {et~al.}(2010)\citenamefont {Xiao},
  \citenamefont {Chang},\ and\ \citenamefont {Niu}}]{Berryphase_Xiao}%
  \BibitemOpen
  \bibfield  {author} {\bibinfo {author} {\bibfnamefont {D.}~\bibnamefont
  {Xiao}}, \bibinfo {author} {\bibfnamefont {M.-C.}\ \bibnamefont {Chang}},\
  and\ \bibinfo {author} {\bibfnamefont {Q.}~\bibnamefont {Niu}},\ }\bibfield
  {title} {\bibinfo {title} {Berry phase effects on electronic properties},\
  }\href {https://doi.org/10.1103/RevModPhys.82.1959} {\bibfield  {journal}
  {\bibinfo  {journal} {Rev. Mod. Phys.}\ }\textbf {\bibinfo {volume} {82}},\
  \bibinfo {pages} {1959} (\bibinfo {year} {2010})}\BibitemShut {NoStop}%
\bibitem [{\citenamefont {Slager}\ \emph {et~al.}(2013)\citenamefont {Slager},
  \citenamefont {Mesaros}, \citenamefont {Juri{\v{c}}i{\'{c}}},\ and\
  \citenamefont {Zaanen}}]{Slager2013}%
  \BibitemOpen
  \bibfield  {author} {\bibinfo {author} {\bibfnamefont {R.-J.}\ \bibnamefont
  {Slager}}, \bibinfo {author} {\bibfnamefont {A.}~\bibnamefont {Mesaros}},
  \bibinfo {author} {\bibfnamefont {V.}~\bibnamefont {Juri{\v{c}}i{\'{c}}}},\
  and\ \bibinfo {author} {\bibfnamefont {J.}~\bibnamefont {Zaanen}},\
  }\bibfield  {title} {\bibinfo {title} {The space group classification of
  topological band-insulators},\ }\href {https://doi.org/10.1038/nphys2513}
  {\bibfield  {journal} {\bibinfo  {journal} {Nature Physics}\ }\textbf
  {\bibinfo {volume} {9}},\ \bibinfo {pages} {98} (\bibinfo {year}
  {2013})}\BibitemShut {NoStop}%
\bibitem [{\citenamefont {Kruthoff}\ \emph {et~al.}(2017)\citenamefont
  {Kruthoff}, \citenamefont {de~Boer}, \citenamefont {van Wezel}, \citenamefont
  {Kane},\ and\ \citenamefont {Slager}}]{kruthoff2017topo}%
  \BibitemOpen
  \bibfield  {author} {\bibinfo {author} {\bibfnamefont {J.}~\bibnamefont
  {Kruthoff}}, \bibinfo {author} {\bibfnamefont {J.}~\bibnamefont {de~Boer}},
  \bibinfo {author} {\bibfnamefont {J.}~\bibnamefont {van Wezel}}, \bibinfo
  {author} {\bibfnamefont {C.~L.}\ \bibnamefont {Kane}},\ and\ \bibinfo
  {author} {\bibfnamefont {R.-J.}\ \bibnamefont {Slager}},\ }\bibfield  {title}
  {\bibinfo {title} {Topological classification of crystalline insulators
  through band structure combinatorics},\ }\href
  {https://doi.org/10.1103/PhysRevX.7.041069} {\bibfield  {journal} {\bibinfo
  {journal} {Phys. Rev. X}\ }\textbf {\bibinfo {volume} {7}},\ \bibinfo {pages}
  {041069} (\bibinfo {year} {2017})}\BibitemShut {NoStop}%
\bibitem [{\citenamefont {Bradlyn}\ \emph {et~al.}(2017)\citenamefont
  {Bradlyn}, \citenamefont {Elcoro}, \citenamefont {Cano}, \citenamefont
  {Vergniory}, \citenamefont {Wang}, \citenamefont {Felser}, \citenamefont
  {Aroyo},\ and\ \citenamefont {Bernevig}}]{TQC_Bradlyn2017}%
  \BibitemOpen
  \bibfield  {author} {\bibinfo {author} {\bibfnamefont {B.}~\bibnamefont
  {Bradlyn}}, \bibinfo {author} {\bibfnamefont {L.}~\bibnamefont {Elcoro}},
  \bibinfo {author} {\bibfnamefont {J.}~\bibnamefont {Cano}}, \bibinfo {author}
  {\bibfnamefont {M.~G.}\ \bibnamefont {Vergniory}}, \bibinfo {author}
  {\bibfnamefont {Z.}~\bibnamefont {Wang}}, \bibinfo {author} {\bibfnamefont
  {C.}~\bibnamefont {Felser}}, \bibinfo {author} {\bibfnamefont {M.~I.}\
  \bibnamefont {Aroyo}},\ and\ \bibinfo {author} {\bibfnamefont {B.~A.}\
  \bibnamefont {Bernevig}},\ }\bibfield  {title} {\bibinfo {title} {Topological
  quantum chemistry},\ }\href {https://doi.org/10.1038/nature23268} {\bibfield
  {journal} {\bibinfo  {journal} {Nature}\ }\textbf {\bibinfo {volume} {547}},\
  \bibinfo {pages} {298} (\bibinfo {year} {2017})}\BibitemShut {NoStop}%
\bibitem [{\citenamefont {Song}\ \emph {et~al.}(2018)\citenamefont {Song},
  \citenamefont {Zhang}, \citenamefont {Fang},\ and\ \citenamefont
  {Fang}}]{SI_Song2018}%
  \BibitemOpen
  \bibfield  {author} {\bibinfo {author} {\bibfnamefont {Z.}~\bibnamefont
  {Song}}, \bibinfo {author} {\bibfnamefont {T.}~\bibnamefont {Zhang}},
  \bibinfo {author} {\bibfnamefont {Z.}~\bibnamefont {Fang}},\ and\ \bibinfo
  {author} {\bibfnamefont {C.}~\bibnamefont {Fang}},\ }\bibfield  {title}
  {\bibinfo {title} {Quantitative mappings between symmetry and topology in
  solids},\ }\href {https://doi.org/10.1038/s41467-018-06010-w} {\bibfield
  {journal} {\bibinfo  {journal} {Nat. Commun.}\ }\textbf {\bibinfo {volume}
  {9}},\ \bibinfo {pages} {3530} (\bibinfo {year} {2018})}\BibitemShut
  {NoStop}%
\bibitem [{\citenamefont {Zhang}\ \emph {et~al.}(2019)\citenamefont {Zhang},
  \citenamefont {Jiang}, \citenamefont {Song}, \citenamefont {Huang},
  \citenamefont {He}, \citenamefont {Fang}, \citenamefont {Weng},\ and\
  \citenamefont {Fang}}]{Database1_CFang}%
  \BibitemOpen
  \bibfield  {author} {\bibinfo {author} {\bibfnamefont {T.}~\bibnamefont
  {Zhang}}, \bibinfo {author} {\bibfnamefont {Y.}~\bibnamefont {Jiang}},
  \bibinfo {author} {\bibfnamefont {Z.}~\bibnamefont {Song}}, \bibinfo {author}
  {\bibfnamefont {H.}~\bibnamefont {Huang}}, \bibinfo {author} {\bibfnamefont
  {Y.}~\bibnamefont {He}}, \bibinfo {author} {\bibfnamefont {Z.}~\bibnamefont
  {Fang}}, \bibinfo {author} {\bibfnamefont {H.}~\bibnamefont {Weng}},\ and\
  \bibinfo {author} {\bibfnamefont {C.}~\bibnamefont {Fang}},\ }\bibfield
  {title} {\bibinfo {title} {Catalogue of topological electronic materials},\
  }\href {https://doi.org/10.1038/s41586-019-0944-6} {\bibfield  {journal}
  {\bibinfo  {journal} {Nature}\ }\textbf {\bibinfo {volume} {566}},\ \bibinfo
  {pages} {475} (\bibinfo {year} {2019})}\BibitemShut {NoStop}%
\bibitem [{\citenamefont {Vergniory}\ \emph {et~al.}(2019)\citenamefont
  {Vergniory}, \citenamefont {Elcoro}, \citenamefont {Felser}, \citenamefont
  {Regnault}, \citenamefont {Bernevig},\ and\ \citenamefont
  {Wang}}]{Database2_AB2019}%
  \BibitemOpen
  \bibfield  {author} {\bibinfo {author} {\bibfnamefont {M.~G.}\ \bibnamefont
  {Vergniory}}, \bibinfo {author} {\bibfnamefont {L.}~\bibnamefont {Elcoro}},
  \bibinfo {author} {\bibfnamefont {C.}~\bibnamefont {Felser}}, \bibinfo
  {author} {\bibfnamefont {N.}~\bibnamefont {Regnault}}, \bibinfo {author}
  {\bibfnamefont {B.~A.}\ \bibnamefont {Bernevig}},\ and\ \bibinfo {author}
  {\bibfnamefont {Z.}~\bibnamefont {Wang}},\ }\bibfield  {title} {\bibinfo
  {title} {A complete catalogue of high-quality topological materials},\ }\href
  {https://doi.org/10.1038/s41586-019-0954-4} {\bibfield  {journal} {\bibinfo
  {journal} {Nature}\ }\textbf {\bibinfo {volume} {566}},\ \bibinfo {pages}
  {480} (\bibinfo {year} {2019})}\BibitemShut {NoStop}%
\bibitem [{\citenamefont {Tang}\ \emph {et~al.}(2019)\citenamefont {Tang},
  \citenamefont {Po}, \citenamefont {Vishwanath},\ and\ \citenamefont
  {Wan}}]{Database3_Ashvin2019}%
  \BibitemOpen
  \bibfield  {author} {\bibinfo {author} {\bibfnamefont {F.}~\bibnamefont
  {Tang}}, \bibinfo {author} {\bibfnamefont {H.~C.}\ \bibnamefont {Po}},
  \bibinfo {author} {\bibfnamefont {A.}~\bibnamefont {Vishwanath}},\ and\
  \bibinfo {author} {\bibfnamefont {X.}~\bibnamefont {Wan}},\ }\bibfield
  {title} {\bibinfo {title} {Comprehensive search for topological materials
  using symmetry indicators},\ }\href
  {https://doi.org/10.1038/s41586-019-0937-5} {\bibfield  {journal} {\bibinfo
  {journal} {Nature}\ }\textbf {\bibinfo {volume} {566}},\ \bibinfo {pages}
  {486} (\bibinfo {year} {2019})}\BibitemShut {NoStop}%
\bibitem [{\citenamefont {Lin}\ \emph {et~al.}(2013)\citenamefont {Lin},
  \citenamefont {Das}, \citenamefont {Okada}, \citenamefont {Boyer},
  \citenamefont {Wise}, \citenamefont {Tomasik}, \citenamefont {Zhen},
  \citenamefont {Hudson}, \citenamefont {Zhou}, \citenamefont {Madhavan},
  \citenamefont {Ren}, \citenamefont {Ikuta},\ and\ \citenamefont
  {Bansil}}]{Dandling_Lin2013}%
  \BibitemOpen
  \bibfield  {author} {\bibinfo {author} {\bibfnamefont {H.}~\bibnamefont
  {Lin}}, \bibinfo {author} {\bibfnamefont {T.}~\bibnamefont {Das}}, \bibinfo
  {author} {\bibfnamefont {Y.}~\bibnamefont {Okada}}, \bibinfo {author}
  {\bibfnamefont {M.~C.}\ \bibnamefont {Boyer}}, \bibinfo {author}
  {\bibfnamefont {W.~D.}\ \bibnamefont {Wise}}, \bibinfo {author}
  {\bibfnamefont {M.}~\bibnamefont {Tomasik}}, \bibinfo {author} {\bibfnamefont
  {B.}~\bibnamefont {Zhen}}, \bibinfo {author} {\bibfnamefont {E.~W.}\
  \bibnamefont {Hudson}}, \bibinfo {author} {\bibfnamefont {W.}~\bibnamefont
  {Zhou}}, \bibinfo {author} {\bibfnamefont {V.}~\bibnamefont {Madhavan}},
  \bibinfo {author} {\bibfnamefont {C.-Y.}\ \bibnamefont {Ren}}, \bibinfo
  {author} {\bibfnamefont {H.}~\bibnamefont {Ikuta}},\ and\ \bibinfo {author}
  {\bibfnamefont {A.}~\bibnamefont {Bansil}},\ }\bibfield  {title} {\bibinfo
  {title} {Topological dangling bonds with large spin splitting and enhanced
  spin polarization on the surfaces of
  ${{\mathrm{Bi}}}_{2}{{\mathrm{Se}}}_{3}$},\ }\href
  {https://doi.org/10.1021/nl304099x} {\bibfield  {journal} {\bibinfo
  {journal} {Nano Lett.}\ }\textbf {\bibinfo {volume} {13}},\ \bibinfo {pages}
  {1915} (\bibinfo {year} {2013})}\BibitemShut {NoStop}%
\bibitem [{\citenamefont {Singh}\ \emph {et~al.}(2016)\citenamefont {Singh},
  \citenamefont {Lin}, \citenamefont {Prasad},\ and\ \citenamefont
  {Bansil}}]{Singh_termination}%
  \BibitemOpen
  \bibfield  {author} {\bibinfo {author} {\bibfnamefont {B.}~\bibnamefont
  {Singh}}, \bibinfo {author} {\bibfnamefont {H.}~\bibnamefont {Lin}}, \bibinfo
  {author} {\bibfnamefont {R.}~\bibnamefont {Prasad}},\ and\ \bibinfo {author}
  {\bibfnamefont {A.}~\bibnamefont {Bansil}},\ }\bibfield  {title} {\bibinfo
  {title} {Role of surface termination in realizing well-isolated topological
  surface states within the bulk band gap in {TlBiSe}$_{2}$ and
  {TlBiTe}$_{2}$},\ }\href {https://doi.org/10.1103/PhysRevB.93.085113}
  {\bibfield  {journal} {\bibinfo  {journal} {Phys. Rev. B}\ }\textbf {\bibinfo
  {volume} {93}},\ \bibinfo {pages} {085113} (\bibinfo {year}
  {2016})}\BibitemShut {NoStop}%
\bibitem [{\citenamefont {Singh}\ \emph
  {et~al.}(2018{\natexlab{a}})\citenamefont {Singh}, \citenamefont {Zhou},
  \citenamefont {Lin},\ and\ \citenamefont {Bansil}}]{singh2018saddle}%
  \BibitemOpen
  \bibfield  {author} {\bibinfo {author} {\bibfnamefont {B.}~\bibnamefont
  {Singh}}, \bibinfo {author} {\bibfnamefont {X.}~\bibnamefont {Zhou}},
  \bibinfo {author} {\bibfnamefont {H.}~\bibnamefont {Lin}},\ and\ \bibinfo
  {author} {\bibfnamefont {A.}~\bibnamefont {Bansil}},\ }\bibfield  {title}
  {\bibinfo {title} {Saddle-like topological surface states on the
  ${T}{T}^{\ensuremath{'}}{X}$ family of compounds (${T}, {T}^{\ensuremath{'}}$
  = transition metal, ${X}=\mathrm{Si}$, ge)},\ }\href
  {https://doi.org/10.1103/PhysRevB.97.075125} {\bibfield  {journal} {\bibinfo
  {journal} {Phys. Rev. B}\ }\textbf {\bibinfo {volume} {97}},\ \bibinfo
  {pages} {075125} (\bibinfo {year} {2018}{\natexlab{a}})}\BibitemShut
  {NoStop}%
\bibitem [{\citenamefont {Burkov}\ \emph {et~al.}(2011)\citenamefont {Burkov},
  \citenamefont {Hook},\ and\ \citenamefont {Balents}}]{NodalLine_Balents}%
  \BibitemOpen
  \bibfield  {author} {\bibinfo {author} {\bibfnamefont {A.~A.}\ \bibnamefont
  {Burkov}}, \bibinfo {author} {\bibfnamefont {M.~D.}\ \bibnamefont {Hook}},\
  and\ \bibinfo {author} {\bibfnamefont {L.}~\bibnamefont {Balents}},\
  }\bibfield  {title} {\bibinfo {title} {Topological nodal semimetals},\ }\href
  {https://doi.org/10.1103/PhysRevB.84.235126} {\bibfield  {journal} {\bibinfo
  {journal} {Phys. Rev. B}\ }\textbf {\bibinfo {volume} {84}},\ \bibinfo
  {pages} {235126} (\bibinfo {year} {2011})}\BibitemShut {NoStop}%
\bibitem [{\citenamefont {Fang}\ \emph {et~al.}(2015)\citenamefont {Fang},
  \citenamefont {Chen}, \citenamefont {Kee},\ and\ \citenamefont
  {Fu}}]{Nodal_Liang}%
  \BibitemOpen
  \bibfield  {author} {\bibinfo {author} {\bibfnamefont {C.}~\bibnamefont
  {Fang}}, \bibinfo {author} {\bibfnamefont {Y.}~\bibnamefont {Chen}}, \bibinfo
  {author} {\bibfnamefont {H.-Y.}\ \bibnamefont {Kee}},\ and\ \bibinfo {author}
  {\bibfnamefont {L.}~\bibnamefont {Fu}},\ }\bibfield  {title} {\bibinfo
  {title} {Topological nodal line semimetals with and without spin-orbital
  coupling},\ }\href {https://doi.org/10.1103/PhysRevB.92.081201} {\bibfield
  {journal} {\bibinfo  {journal} {Phys. Rev. B}\ }\textbf {\bibinfo {volume}
  {92}},\ \bibinfo {pages} {081201} (\bibinfo {year} {2015})}\BibitemShut
  {NoStop}%
\bibitem [{\citenamefont {Bian}\ \emph {et~al.}(2016)\citenamefont {Bian},
  \citenamefont {Chang}, \citenamefont {Zheng}, \citenamefont {Velury},
  \citenamefont {Xu}, \citenamefont {Neupert}, \citenamefont {Chiu},
  \citenamefont {Huang}, \citenamefont {Sanchez}, \citenamefont {Belopolski},
  \citenamefont {Alidoust}, \citenamefont {Chen}, \citenamefont {Chang},
  \citenamefont {Bansil}, \citenamefont {Jeng}, \citenamefont {Lin},\ and\
  \citenamefont {Hasan}}]{bian2016drumhead}%
  \BibitemOpen
  \bibfield  {author} {\bibinfo {author} {\bibfnamefont {G.}~\bibnamefont
  {Bian}}, \bibinfo {author} {\bibfnamefont {T.-R.}\ \bibnamefont {Chang}},
  \bibinfo {author} {\bibfnamefont {H.}~\bibnamefont {Zheng}}, \bibinfo
  {author} {\bibfnamefont {S.}~\bibnamefont {Velury}}, \bibinfo {author}
  {\bibfnamefont {S.-Y.}\ \bibnamefont {Xu}}, \bibinfo {author} {\bibfnamefont
  {T.}~\bibnamefont {Neupert}}, \bibinfo {author} {\bibfnamefont {C.-K.}\
  \bibnamefont {Chiu}}, \bibinfo {author} {\bibfnamefont {S.-M.}\ \bibnamefont
  {Huang}}, \bibinfo {author} {\bibfnamefont {D.~S.}\ \bibnamefont {Sanchez}},
  \bibinfo {author} {\bibfnamefont {I.}~\bibnamefont {Belopolski}}, \bibinfo
  {author} {\bibfnamefont {N.}~\bibnamefont {Alidoust}}, \bibinfo {author}
  {\bibfnamefont {P.-J.}\ \bibnamefont {Chen}}, \bibinfo {author}
  {\bibfnamefont {G.}~\bibnamefont {Chang}}, \bibinfo {author} {\bibfnamefont
  {A.}~\bibnamefont {Bansil}}, \bibinfo {author} {\bibfnamefont {H.-T.}\
  \bibnamefont {Jeng}}, \bibinfo {author} {\bibfnamefont {H.}~\bibnamefont
  {Lin}},\ and\ \bibinfo {author} {\bibfnamefont {M.~Z.}\ \bibnamefont
  {Hasan}},\ }\bibfield  {title} {\bibinfo {title} {Drumhead surface states and
  topological nodal-line fermions in {TlTaSe}$_{2}$},\ }\href
  {https://doi.org/10.1103/PhysRevB.93.121113} {\bibfield  {journal} {\bibinfo
  {journal} {Phys. Rev. B}\ }\textbf {\bibinfo {volume} {93}},\ \bibinfo
  {pages} {121113} (\bibinfo {year} {2016})}\BibitemShut {NoStop}%
\bibitem [{\citenamefont {Chan}\ \emph {et~al.}(2016)\citenamefont {Chan},
  \citenamefont {Chiu}, \citenamefont {Chou},\ and\ \citenamefont
  {Schnyder}}]{chan2016ca}%
  \BibitemOpen
  \bibfield  {author} {\bibinfo {author} {\bibfnamefont {Y.-H.}\ \bibnamefont
  {Chan}}, \bibinfo {author} {\bibfnamefont {C.-K.}\ \bibnamefont {Chiu}},
  \bibinfo {author} {\bibfnamefont {M.~Y.}\ \bibnamefont {Chou}},\ and\
  \bibinfo {author} {\bibfnamefont {A.~P.}\ \bibnamefont {Schnyder}},\
  }\bibfield  {title} {\bibinfo {title}
  {${{\mathrm{Ca}}}_{3}{{\mathrm{P}}}_{2}$ and other topological semimetals
  with line nodes and drumhead surface states},\ }\href
  {https://doi.org/10.1103/PhysRevB.93.205132} {\bibfield  {journal} {\bibinfo
  {journal} {Phys. Rev. B}\ }\textbf {\bibinfo {volume} {93}},\ \bibinfo
  {pages} {205132} (\bibinfo {year} {2016})}\BibitemShut {NoStop}%
\bibitem [{\citenamefont {Singh}\ \emph
  {et~al.}(2018{\natexlab{b}})\citenamefont {Singh}, \citenamefont {Mardanya},
  \citenamefont {Su}, \citenamefont {Lin}, \citenamefont {Agarwal},\ and\
  \citenamefont {Bansil}}]{StarFruit_SM}%
  \BibitemOpen
  \bibfield  {author} {\bibinfo {author} {\bibfnamefont {B.}~\bibnamefont
  {Singh}}, \bibinfo {author} {\bibfnamefont {S.}~\bibnamefont {Mardanya}},
  \bibinfo {author} {\bibfnamefont {C.}~\bibnamefont {Su}}, \bibinfo {author}
  {\bibfnamefont {H.}~\bibnamefont {Lin}}, \bibinfo {author} {\bibfnamefont
  {A.}~\bibnamefont {Agarwal}},\ and\ \bibinfo {author} {\bibfnamefont
  {A.}~\bibnamefont {Bansil}},\ }\bibfield  {title} {\bibinfo {title}
  {Spin-orbit coupling driven crossover from a starfruitlike nodal semimetal to
  dirac and weyl semimetal state in {CaAuAs}},\ }\href
  {https://doi.org/10.1103/PhysRevB.98.085122} {\bibfield  {journal} {\bibinfo
  {journal} {Phys. Rev. B}\ }\textbf {\bibinfo {volume} {98}},\ \bibinfo
  {pages} {085122} (\bibinfo {year} {2018}{\natexlab{b}})}\BibitemShut
  {NoStop}%
\bibitem [{\citenamefont {Xu}\ \emph {et~al.}(2017)\citenamefont {Xu},
  \citenamefont {Alidoust}, \citenamefont {Chang}, \citenamefont {Lu},
  \citenamefont {Singh}, \citenamefont {Belopolski}, \citenamefont {Sanchez},
  \citenamefont {Zhang}, \citenamefont {Bian}, \citenamefont {Zheng},
  \citenamefont {Husanu}, \citenamefont {Bian}, \citenamefont {Huang},
  \citenamefont {Hsu}, \citenamefont {Chang}, \citenamefont {Jeng},
  \citenamefont {Bansil}, \citenamefont {Neupert}, \citenamefont {Strocov},
  \citenamefont {Lin}, \citenamefont {Jia},\ and\ \citenamefont
  {Hasan}}]{TypeII_LaAlGe}%
  \BibitemOpen
  \bibfield  {author} {\bibinfo {author} {\bibfnamefont {S.-Y.}\ \bibnamefont
  {Xu}}, \bibinfo {author} {\bibfnamefont {N.}~\bibnamefont {Alidoust}},
  \bibinfo {author} {\bibfnamefont {G.}~\bibnamefont {Chang}}, \bibinfo
  {author} {\bibfnamefont {H.}~\bibnamefont {Lu}}, \bibinfo {author}
  {\bibfnamefont {B.}~\bibnamefont {Singh}}, \bibinfo {author} {\bibfnamefont
  {I.}~\bibnamefont {Belopolski}}, \bibinfo {author} {\bibfnamefont {D.~S.}\
  \bibnamefont {Sanchez}}, \bibinfo {author} {\bibfnamefont {X.}~\bibnamefont
  {Zhang}}, \bibinfo {author} {\bibfnamefont {G.}~\bibnamefont {Bian}},
  \bibinfo {author} {\bibfnamefont {H.}~\bibnamefont {Zheng}}, \bibinfo
  {author} {\bibfnamefont {M.-A.}\ \bibnamefont {Husanu}}, \bibinfo {author}
  {\bibfnamefont {Y.}~\bibnamefont {Bian}}, \bibinfo {author} {\bibfnamefont
  {S.-M.}\ \bibnamefont {Huang}}, \bibinfo {author} {\bibfnamefont {C.-H.}\
  \bibnamefont {Hsu}}, \bibinfo {author} {\bibfnamefont {T.-R.}\ \bibnamefont
  {Chang}}, \bibinfo {author} {\bibfnamefont {H.-T.}\ \bibnamefont {Jeng}},
  \bibinfo {author} {\bibfnamefont {A.}~\bibnamefont {Bansil}}, \bibinfo
  {author} {\bibfnamefont {T.}~\bibnamefont {Neupert}}, \bibinfo {author}
  {\bibfnamefont {V.~N.}\ \bibnamefont {Strocov}}, \bibinfo {author}
  {\bibfnamefont {H.}~\bibnamefont {Lin}}, \bibinfo {author} {\bibfnamefont
  {S.}~\bibnamefont {Jia}},\ and\ \bibinfo {author} {\bibfnamefont {M.~Z.}\
  \bibnamefont {Hasan}},\ }\bibfield  {title} {\bibinfo {title} {Discovery of
  lorentz-violating type ii weyl fermions in {LaAlGe}},\ }\href
  {https://doi.org/10.1126/sciadv.1603266} {\bibfield  {journal} {\bibinfo
  {journal} {Sci. Adv.}\ }\textbf {\bibinfo {volume} {3}},\ \bibinfo {pages}
  {e1603266} (\bibinfo {year} {2017})}\BibitemShut {NoStop}%
\bibitem [{\citenamefont {Chang}\ \emph {et~al.}(2019)\citenamefont {Chang},
  \citenamefont {Pletikosic}, \citenamefont {Kong}, \citenamefont {Bian},
  \citenamefont {Huang}, \citenamefont {Denlinger}, \citenamefont {Kushwaha},
  \citenamefont {Sinkovic}, \citenamefont {Jeng}, \citenamefont {Valla},
  \citenamefont {Xie},\ and\ \citenamefont {Cava}}]{chang2019realization}%
  \BibitemOpen
  \bibfield  {author} {\bibinfo {author} {\bibfnamefont {T.-R.}\ \bibnamefont
  {Chang}}, \bibinfo {author} {\bibfnamefont {I.}~\bibnamefont {Pletikosic}},
  \bibinfo {author} {\bibfnamefont {T.}~\bibnamefont {Kong}}, \bibinfo {author}
  {\bibfnamefont {G.}~\bibnamefont {Bian}}, \bibinfo {author} {\bibfnamefont
  {A.}~\bibnamefont {Huang}}, \bibinfo {author} {\bibfnamefont
  {J.}~\bibnamefont {Denlinger}}, \bibinfo {author} {\bibfnamefont {S.~K.}\
  \bibnamefont {Kushwaha}}, \bibinfo {author} {\bibfnamefont {B.}~\bibnamefont
  {Sinkovic}}, \bibinfo {author} {\bibfnamefont {H.-T.}\ \bibnamefont {Jeng}},
  \bibinfo {author} {\bibfnamefont {T.}~\bibnamefont {Valla}}, \bibinfo
  {author} {\bibfnamefont {W.}~\bibnamefont {Xie}},\ and\ \bibinfo {author}
  {\bibfnamefont {R.~J.}\ \bibnamefont {Cava}},\ }\bibfield  {title} {\bibinfo
  {title} {Realization of a type-ii nodal-line semimetal in
  ${{\mathrm{Mg}}}_3{{\mathrm {Bi}}}_2$},\ }\href
  {https://doi.org/https://doi.org/10.1002/advs.201800897} {\bibfield
  {journal} {\bibinfo  {journal} {Adv. Sci.}\ }\textbf {\bibinfo {volume}
  {6}},\ \bibinfo {pages} {1800897} (\bibinfo {year} {2019})}\BibitemShut
  {NoStop}%
\bibitem [{\citenamefont {Wang}\ \emph {et~al.}(2019)\citenamefont {Wang},
  \citenamefont {Singh}, \citenamefont {Ghosh}, \citenamefont {Chiu},
  \citenamefont {Hosen}, \citenamefont {Zhang}, \citenamefont {Ying},
  \citenamefont {Neupane}, \citenamefont {Agarwal}, \citenamefont {Lin},\ and\
  \citenamefont {Bansil}}]{wang2019topo}%
  \BibitemOpen
  \bibfield  {author} {\bibinfo {author} {\bibfnamefont {B.}~\bibnamefont
  {Wang}}, \bibinfo {author} {\bibfnamefont {B.}~\bibnamefont {Singh}},
  \bibinfo {author} {\bibfnamefont {B.}~\bibnamefont {Ghosh}}, \bibinfo
  {author} {\bibfnamefont {W.-C.}\ \bibnamefont {Chiu}}, \bibinfo {author}
  {\bibfnamefont {M.~M.}\ \bibnamefont {Hosen}}, \bibinfo {author}
  {\bibfnamefont {Q.}~\bibnamefont {Zhang}}, \bibinfo {author} {\bibfnamefont
  {L.}~\bibnamefont {Ying}}, \bibinfo {author} {\bibfnamefont {M.}~\bibnamefont
  {Neupane}}, \bibinfo {author} {\bibfnamefont {A.}~\bibnamefont {Agarwal}},
  \bibinfo {author} {\bibfnamefont {H.}~\bibnamefont {Lin}},\ and\ \bibinfo
  {author} {\bibfnamefont {A.}~\bibnamefont {Bansil}},\ }\bibfield  {title}
  {\bibinfo {title} {Topological crystalline insulator state with type-ii dirac
  fermions in transition metal dipnictides},\ }\href
  {https://doi.org/10.1103/PhysRevB.100.205118} {\bibfield  {journal} {\bibinfo
   {journal} {Phys. Rev. B}\ }\textbf {\bibinfo {volume} {100}},\ \bibinfo
  {pages} {205118} (\bibinfo {year} {2019})}\BibitemShut {NoStop}%
\bibitem [{\citenamefont {Li}\ \emph {et~al.}(2016)\citenamefont {Li},
  \citenamefont {Luo}, \citenamefont {Dai}, \citenamefont {Yu}, \citenamefont
  {Zhang},\ and\ \citenamefont {Chen}}]{hybrid_Weyl}%
  \BibitemOpen
  \bibfield  {author} {\bibinfo {author} {\bibfnamefont {F.-Y.}\ \bibnamefont
  {Li}}, \bibinfo {author} {\bibfnamefont {X.}~\bibnamefont {Luo}}, \bibinfo
  {author} {\bibfnamefont {X.}~\bibnamefont {Dai}}, \bibinfo {author}
  {\bibfnamefont {Y.}~\bibnamefont {Yu}}, \bibinfo {author} {\bibfnamefont
  {F.}~\bibnamefont {Zhang}},\ and\ \bibinfo {author} {\bibfnamefont
  {G.}~\bibnamefont {Chen}},\ }\bibfield  {title} {\bibinfo {title} {Hybrid
  weyl semimetal},\ }\href {https://doi.org/10.1103/PhysRevB.94.121105}
  {\bibfield  {journal} {\bibinfo  {journal} {Phys. Rev. B}\ }\textbf {\bibinfo
  {volume} {94}},\ \bibinfo {pages} {121105} (\bibinfo {year}
  {2016})}\BibitemShut {NoStop}%
\bibitem [{\citenamefont {Zhang}\ \emph {et~al.}(2018)\citenamefont {Zhang},
  \citenamefont {Yu}, \citenamefont {Lu}, \citenamefont {Sheng}, \citenamefont
  {Yang},\ and\ \citenamefont {Yang}}]{hybrid_zhang2018}%
  \BibitemOpen
  \bibfield  {author} {\bibinfo {author} {\bibfnamefont {X.}~\bibnamefont
  {Zhang}}, \bibinfo {author} {\bibfnamefont {Z.-M.}\ \bibnamefont {Yu}},
  \bibinfo {author} {\bibfnamefont {Y.}~\bibnamefont {Lu}}, \bibinfo {author}
  {\bibfnamefont {X.-L.}\ \bibnamefont {Sheng}}, \bibinfo {author}
  {\bibfnamefont {H.~Y.}\ \bibnamefont {Yang}},\ and\ \bibinfo {author}
  {\bibfnamefont {S.~A.}\ \bibnamefont {Yang}},\ }\bibfield  {title} {\bibinfo
  {title} {Hybrid nodal loop metal: Unconventional magnetoresponse and material
  realization},\ }\href {https://doi.org/10.1103/PhysRevB.97.125143} {\bibfield
   {journal} {\bibinfo  {journal} {Phys. Rev. B}\ }\textbf {\bibinfo {volume}
  {97}},\ \bibinfo {pages} {125143} (\bibinfo {year} {2018})}\BibitemShut
  {NoStop}%
\bibitem [{\citenamefont {Shao}\ \emph {et~al.}(2020)\citenamefont {Shao},
  \citenamefont {Rudenko}, \citenamefont {Hu}, \citenamefont {Sun},
  \citenamefont {Zhu}, \citenamefont {Moon}, \citenamefont {Millis},
  \citenamefont {Yuan}, \citenamefont {Lichtenstein}, \citenamefont {Smirnov},
  \citenamefont {Mao}, \citenamefont {Katsnelson},\ and\ \citenamefont
  {Basov}}]{Correlation_Shao2020}%
  \BibitemOpen
  \bibfield  {author} {\bibinfo {author} {\bibfnamefont {Y.}~\bibnamefont
  {Shao}}, \bibinfo {author} {\bibfnamefont {A.~N.}\ \bibnamefont {Rudenko}},
  \bibinfo {author} {\bibfnamefont {J.}~\bibnamefont {Hu}}, \bibinfo {author}
  {\bibfnamefont {Z.}~\bibnamefont {Sun}}, \bibinfo {author} {\bibfnamefont
  {Y.}~\bibnamefont {Zhu}}, \bibinfo {author} {\bibfnamefont {S.}~\bibnamefont
  {Moon}}, \bibinfo {author} {\bibfnamefont {A.~J.}\ \bibnamefont {Millis}},
  \bibinfo {author} {\bibfnamefont {S.}~\bibnamefont {Yuan}}, \bibinfo {author}
  {\bibfnamefont {A.~I.}\ \bibnamefont {Lichtenstein}}, \bibinfo {author}
  {\bibfnamefont {D.}~\bibnamefont {Smirnov}}, \bibinfo {author} {\bibfnamefont
  {Z.~Q.}\ \bibnamefont {Mao}}, \bibinfo {author} {\bibfnamefont {M.~I.}\
  \bibnamefont {Katsnelson}},\ and\ \bibinfo {author} {\bibfnamefont {D.~N.}\
  \bibnamefont {Basov}},\ }\bibfield  {title} {\bibinfo {title} {Electronic
  correlations in nodal-line semimetals},\ }\href
  {https://doi.org/10.1038/s41567-020-0859-z} {\bibfield  {journal} {\bibinfo
  {journal} {Nat. Phys.}\ }\textbf {\bibinfo {volume} {16}},\ \bibinfo {pages}
  {636} (\bibinfo {year} {2020})}\BibitemShut {NoStop}%
\bibitem [{\citenamefont {Kou}\ \emph {et~al.}(2015)\citenamefont {Kou},
  \citenamefont {Chen},\ and\ \citenamefont {Smith}}]{BP_Review_Kou2015}%
  \BibitemOpen
  \bibfield  {author} {\bibinfo {author} {\bibfnamefont {L.}~\bibnamefont
  {Kou}}, \bibinfo {author} {\bibfnamefont {C.}~\bibnamefont {Chen}},\ and\
  \bibinfo {author} {\bibfnamefont {S.~C.}\ \bibnamefont {Smith}},\ }\bibfield
  {title} {\bibinfo {title} {Phosphorene: Fabrication, properties, and
  applications},\ }\href {https://doi.org/10.1021/acs.jpclett.5b01094}
  {\bibfield  {journal} {\bibinfo  {journal} {J. Phys. Chem. Lett.}\ }\textbf
  {\bibinfo {volume} {6}},\ \bibinfo {pages} {2794} (\bibinfo {year}
  {2015})}\BibitemShut {NoStop}%
\bibitem [{\citenamefont {Xia}\ \emph {et~al.}(2019)\citenamefont {Xia},
  \citenamefont {Wang}, \citenamefont {Hwang}, \citenamefont {Neto},\ and\
  \citenamefont {Yang}}]{xia2019black}%
  \BibitemOpen
  \bibfield  {author} {\bibinfo {author} {\bibfnamefont {F.}~\bibnamefont
  {Xia}}, \bibinfo {author} {\bibfnamefont {H.}~\bibnamefont {Wang}}, \bibinfo
  {author} {\bibfnamefont {J.~C.~M.}\ \bibnamefont {Hwang}}, \bibinfo {author}
  {\bibfnamefont {A.~H.~C.}\ \bibnamefont {Neto}},\ and\ \bibinfo {author}
  {\bibfnamefont {L.}~\bibnamefont {Yang}},\ }\bibfield  {title} {\bibinfo
  {title} {Black phosphorus and its isoelectronic materials},\ }\href
  {https://doi.org/10.1038/s42254-019-0043-5} {\bibfield  {journal} {\bibinfo
  {journal} {Nat. Rev. Phys.}\ }\textbf {\bibinfo {volume} {1}},\ \bibinfo
  {pages} {306} (\bibinfo {year} {2019})}\BibitemShut {NoStop}%
\bibitem [{\citenamefont {Qiao}\ \emph {et~al.}(2014)\citenamefont {Qiao},
  \citenamefont {Kong}, \citenamefont {Hu}, \citenamefont {Yang},\ and\
  \citenamefont {Ji}}]{BP_Qiao2014}%
  \BibitemOpen
  \bibfield  {author} {\bibinfo {author} {\bibfnamefont {J.}~\bibnamefont
  {Qiao}}, \bibinfo {author} {\bibfnamefont {X.}~\bibnamefont {Kong}}, \bibinfo
  {author} {\bibfnamefont {Z.-X.}\ \bibnamefont {Hu}}, \bibinfo {author}
  {\bibfnamefont {F.}~\bibnamefont {Yang}},\ and\ \bibinfo {author}
  {\bibfnamefont {W.}~\bibnamefont {Ji}},\ }\bibfield  {title} {\bibinfo
  {title} {High-mobility transport anisotropy and linear dichroism in few-layer
  black phosphorus},\ }\href {https://doi.org/10.1038/ncomms5475} {\bibfield
  {journal} {\bibinfo  {journal} {Nat. Commun.}\ }\textbf {\bibinfo {volume}
  {5}},\ \bibinfo {pages} {4475} (\bibinfo {year} {2014})}\BibitemShut
  {NoStop}%
\bibitem [{\citenamefont {Xia}\ \emph {et~al.}(2014)\citenamefont {Xia},
  \citenamefont {Wang},\ and\ \citenamefont {Jia}}]{BP_Xia2014}%
  \BibitemOpen
  \bibfield  {author} {\bibinfo {author} {\bibfnamefont {F.}~\bibnamefont
  {Xia}}, \bibinfo {author} {\bibfnamefont {H.}~\bibnamefont {Wang}},\ and\
  \bibinfo {author} {\bibfnamefont {Y.}~\bibnamefont {Jia}},\ }\bibfield
  {title} {\bibinfo {title} {Rediscovering black phosphorus as an anisotropic
  layered material for optoelectronics and electronics},\ }\href
  {https://doi.org/10.1038/ncomms5458} {\bibfield  {journal} {\bibinfo
  {journal} {Nat. Commun.}\ }\textbf {\bibinfo {volume} {5}},\ \bibinfo {pages}
  {4458} (\bibinfo {year} {2014})}\BibitemShut {NoStop}%
\bibitem [{\citenamefont {Li}\ \emph {et~al.}(2017)\citenamefont {Li},
  \citenamefont {Tao}, \citenamefont {Chen}, \citenamefont {Fang},
  \citenamefont {Li}, \citenamefont {Wang}, \citenamefont {Xu},\ and\
  \citenamefont {Zhu}}]{SPR_2DXinming2017}%
  \BibitemOpen
  \bibfield  {author} {\bibinfo {author} {\bibfnamefont {X.}~\bibnamefont
  {Li}}, \bibinfo {author} {\bibfnamefont {L.}~\bibnamefont {Tao}}, \bibinfo
  {author} {\bibfnamefont {Z.}~\bibnamefont {Chen}}, \bibinfo {author}
  {\bibfnamefont {H.}~\bibnamefont {Fang}}, \bibinfo {author} {\bibfnamefont
  {X.}~\bibnamefont {Li}}, \bibinfo {author} {\bibfnamefont {X.}~\bibnamefont
  {Wang}}, \bibinfo {author} {\bibfnamefont {J.-B.}\ \bibnamefont {Xu}},\ and\
  \bibinfo {author} {\bibfnamefont {H.}~\bibnamefont {Zhu}},\ }\bibfield
  {title} {\bibinfo {title} {{Graphene and related two-dimensional materials:
  Structure-property relationships for electronics and optoelectronics}},\
  }\href {https://doi.org/10.1063/1.4983646} {\bibfield  {journal} {\bibinfo
  {journal} {Appl. Phys. Rev.}\ }\textbf {\bibinfo {volume} {4}},\ \bibinfo
  {pages} {021306} (\bibinfo {year} {2017})}\BibitemShut {NoStop}%
\bibitem [{\citenamefont {Xiang}\ \emph {et~al.}(2015)\citenamefont {Xiang},
  \citenamefont {Ye}, \citenamefont {Shang}, \citenamefont {Lei}, \citenamefont
  {Wang}, \citenamefont {Yang}, \citenamefont {Liu}, \citenamefont {Meng},
  \citenamefont {Luo}, \citenamefont {Zou}, \citenamefont {Sun}, \citenamefont
  {Zhang},\ and\ \citenamefont {Chen}}]{xiang2015pressure}%
  \BibitemOpen
  \bibfield  {author} {\bibinfo {author} {\bibfnamefont {Z.~J.}\ \bibnamefont
  {Xiang}}, \bibinfo {author} {\bibfnamefont {G.~J.}\ \bibnamefont {Ye}},
  \bibinfo {author} {\bibfnamefont {C.}~\bibnamefont {Shang}}, \bibinfo
  {author} {\bibfnamefont {B.}~\bibnamefont {Lei}}, \bibinfo {author}
  {\bibfnamefont {N.~Z.}\ \bibnamefont {Wang}}, \bibinfo {author}
  {\bibfnamefont {K.~S.}\ \bibnamefont {Yang}}, \bibinfo {author}
  {\bibfnamefont {D.~Y.}\ \bibnamefont {Liu}}, \bibinfo {author} {\bibfnamefont
  {F.~B.}\ \bibnamefont {Meng}}, \bibinfo {author} {\bibfnamefont {X.~G.}\
  \bibnamefont {Luo}}, \bibinfo {author} {\bibfnamefont {L.~J.}\ \bibnamefont
  {Zou}}, \bibinfo {author} {\bibfnamefont {Z.}~\bibnamefont {Sun}}, \bibinfo
  {author} {\bibfnamefont {Y.}~\bibnamefont {Zhang}},\ and\ \bibinfo {author}
  {\bibfnamefont {X.~H.}\ \bibnamefont {Chen}},\ }\bibfield  {title} {\bibinfo
  {title} {Pressure-induced electronic transition in black phosphorus},\ }\href
  {https://doi.org/10.1103/PhysRevLett.115.186403} {\bibfield  {journal}
  {\bibinfo  {journal} {Phys. Rev. Lett.}\ }\textbf {\bibinfo {volume} {115}},\
  \bibinfo {pages} {186403} (\bibinfo {year} {2015})}\BibitemShut {NoStop}%
\bibitem [{\citenamefont {Kim}\ \emph {et~al.}(2015)\citenamefont {Kim},
  \citenamefont {Baik}, \citenamefont {Ryu}, \citenamefont {Sohn},
  \citenamefont {Park}, \citenamefont {Park}, \citenamefont {Denlinger},
  \citenamefont {Yi}, \citenamefont {Choi},\ and\ \citenamefont
  {Kim}}]{Phosphorene_DSM}%
  \BibitemOpen
  \bibfield  {author} {\bibinfo {author} {\bibfnamefont {J.}~\bibnamefont
  {Kim}}, \bibinfo {author} {\bibfnamefont {S.~S.}\ \bibnamefont {Baik}},
  \bibinfo {author} {\bibfnamefont {S.~H.}\ \bibnamefont {Ryu}}, \bibinfo
  {author} {\bibfnamefont {Y.}~\bibnamefont {Sohn}}, \bibinfo {author}
  {\bibfnamefont {S.}~\bibnamefont {Park}}, \bibinfo {author} {\bibfnamefont
  {B.-G.}\ \bibnamefont {Park}}, \bibinfo {author} {\bibfnamefont
  {J.}~\bibnamefont {Denlinger}}, \bibinfo {author} {\bibfnamefont
  {Y.}~\bibnamefont {Yi}}, \bibinfo {author} {\bibfnamefont {H.~J.}\
  \bibnamefont {Choi}},\ and\ \bibinfo {author} {\bibfnamefont {K.~S.}\
  \bibnamefont {Kim}},\ }\bibfield  {title} {\bibinfo {title} {Observation of
  tunable band gap and anisotropic dirac semimetal state in black phosphorus},\
  }\href {https://doi.org/10.1126/science.aaa6486} {\bibfield  {journal}
  {\bibinfo  {journal} {Science}\ }\textbf {\bibinfo {volume} {349}},\ \bibinfo
  {pages} {723} (\bibinfo {year} {2015})}\BibitemShut {NoStop}%
\bibitem [{\citenamefont {Liu}\ \emph {et~al.}(2017)\citenamefont {Liu},
  \citenamefont {Qiu}, \citenamefont {Carvalho}, \citenamefont {Bao},
  \citenamefont {Xu}, \citenamefont {Tan}, \citenamefont {Liu}, \citenamefont
  {Castro~Neto}, \citenamefont {Loh},\ and\ \citenamefont
  {Lu}}]{Phosphorene_Efield}%
  \BibitemOpen
  \bibfield  {author} {\bibinfo {author} {\bibfnamefont {Y.}~\bibnamefont
  {Liu}}, \bibinfo {author} {\bibfnamefont {Z.}~\bibnamefont {Qiu}}, \bibinfo
  {author} {\bibfnamefont {A.}~\bibnamefont {Carvalho}}, \bibinfo {author}
  {\bibfnamefont {Y.}~\bibnamefont {Bao}}, \bibinfo {author} {\bibfnamefont
  {H.}~\bibnamefont {Xu}}, \bibinfo {author} {\bibfnamefont {S.~J.~R.}\
  \bibnamefont {Tan}}, \bibinfo {author} {\bibfnamefont {W.}~\bibnamefont
  {Liu}}, \bibinfo {author} {\bibfnamefont {A.~H.}\ \bibnamefont
  {Castro~Neto}}, \bibinfo {author} {\bibfnamefont {K.~P.}\ \bibnamefont
  {Loh}},\ and\ \bibinfo {author} {\bibfnamefont {J.}~\bibnamefont {Lu}},\
  }\bibfield  {title} {\bibinfo {title} {Gate-tunable giant stark effect in
  few-layer black phosphorus},\ }\href
  {https://doi.org/10.1021/acs.nanolett.6b05381} {\bibfield  {journal}
  {\bibinfo  {journal} {Nano Lett.}\ }\textbf {\bibinfo {volume} {17}},\
  \bibinfo {pages} {1970} (\bibinfo {year} {2017})}\BibitemShut {NoStop}%
\bibitem [{\citenamefont {Ghosh}\ \emph {et~al.}(2016)\citenamefont {Ghosh},
  \citenamefont {Singh}, \citenamefont {Prasad},\ and\ \citenamefont
  {Agarwal}}]{ghosh2016electric}%
  \BibitemOpen
  \bibfield  {author} {\bibinfo {author} {\bibfnamefont {B.}~\bibnamefont
  {Ghosh}}, \bibinfo {author} {\bibfnamefont {B.}~\bibnamefont {Singh}},
  \bibinfo {author} {\bibfnamefont {R.}~\bibnamefont {Prasad}},\ and\ \bibinfo
  {author} {\bibfnamefont {A.}~\bibnamefont {Agarwal}},\ }\bibfield  {title}
  {\bibinfo {title} {Electric-field tunable dirac semimetal state in
  phosphorene thin films},\ }\href {https://doi.org/10.1103/PhysRevB.94.205426}
  {\bibfield  {journal} {\bibinfo  {journal} {Phys. Rev. B}\ }\textbf {\bibinfo
  {volume} {94}},\ \bibinfo {pages} {205426} (\bibinfo {year}
  {2016})}\BibitemShut {NoStop}%
\bibitem [{\citenamefont {Christiansson}\ \emph {et~al.}(2022)\citenamefont
  {Christiansson}, \citenamefont {Petocchi},\ and\ \citenamefont
  {Werner}}]{chris2022superconductivity}%
  \BibitemOpen
  \bibfield  {author} {\bibinfo {author} {\bibfnamefont {V.}~\bibnamefont
  {Christiansson}}, \bibinfo {author} {\bibfnamefont {F.}~\bibnamefont
  {Petocchi}},\ and\ \bibinfo {author} {\bibfnamefont {P.}~\bibnamefont
  {Werner}},\ }\bibfield  {title} {\bibinfo {title} {Superconductivity in black
  phosphorus and the role of dynamical screening},\ }\href
  {https://doi.org/10.1103/PhysRevB.105.174513} {\bibfield  {journal} {\bibinfo
   {journal} {Phys. Rev. B}\ }\textbf {\bibinfo {volume} {105}},\ \bibinfo
  {pages} {174513} (\bibinfo {year} {2022})}\BibitemShut {NoStop}%
\bibitem [{\citenamefont {Gong}\ \emph {et~al.}(2018)\citenamefont {Gong},
  \citenamefont {Deng}, \citenamefont {Wu}, \citenamefont {Wan}, \citenamefont
  {Wang}, \citenamefont {Li}, \citenamefont {Gou},\ and\ \citenamefont
  {Gao}}]{TMBP_Gong2018}%
  \BibitemOpen
  \bibfield  {author} {\bibinfo {author} {\bibfnamefont {N.}~\bibnamefont
  {Gong}}, \bibinfo {author} {\bibfnamefont {C.}~\bibnamefont {Deng}}, \bibinfo
  {author} {\bibfnamefont {L.}~\bibnamefont {Wu}}, \bibinfo {author}
  {\bibfnamefont {B.}~\bibnamefont {Wan}}, \bibinfo {author} {\bibfnamefont
  {Z.}~\bibnamefont {Wang}}, \bibinfo {author} {\bibfnamefont {Z.}~\bibnamefont
  {Li}}, \bibinfo {author} {\bibfnamefont {H.}~\bibnamefont {Gou}},\ and\
  \bibinfo {author} {\bibfnamefont {F.}~\bibnamefont {Gao}},\ }\bibfield
  {title} {\bibinfo {title} {Structural diversity and electronic properties of
  3d transition metal tetraphosphides, $\mathrm{{TMP}}_4$ ($\mathrm{{TM}} =
  \mathrm{{V}}, \mathrm{{Cr}}, \mathrm{{Mn}}$, and $\mathrm{{Fe}}$)},\ }\href
  {https://doi.org/10.1021/acs.inorgchem.8b01380} {\bibfield  {journal}
  {\bibinfo  {journal} {Inorg. Chem.}\ }\textbf {\bibinfo {volume} {57}},\
  \bibinfo {pages} {9385} (\bibinfo {year} {2018})}\BibitemShut {NoStop}%
\bibitem [{\citenamefont {Sun}\ \emph {et~al.}(2015)\citenamefont {Sun},
  \citenamefont {Lee}, \citenamefont {Pasta}, \citenamefont {Yuan},
  \citenamefont {Zheng}, \citenamefont {Sun}, \citenamefont {Li},\ and\
  \citenamefont {Cui}}]{BPNa_Sun2015}%
  \BibitemOpen
  \bibfield  {author} {\bibinfo {author} {\bibfnamefont {J.}~\bibnamefont
  {Sun}}, \bibinfo {author} {\bibfnamefont {H.-W.}\ \bibnamefont {Lee}},
  \bibinfo {author} {\bibfnamefont {M.}~\bibnamefont {Pasta}}, \bibinfo
  {author} {\bibfnamefont {H.}~\bibnamefont {Yuan}}, \bibinfo {author}
  {\bibfnamefont {G.}~\bibnamefont {Zheng}}, \bibinfo {author} {\bibfnamefont
  {Y.}~\bibnamefont {Sun}}, \bibinfo {author} {\bibfnamefont {Y.}~\bibnamefont
  {Li}},\ and\ \bibinfo {author} {\bibfnamefont {Y.}~\bibnamefont {Cui}},\
  }\bibfield  {title} {\bibinfo {title} {A phosphorene--graphene hybrid
  material as a high-capacity anode for sodium-ion batteries},\ }\href
  {https://doi.org/10.1038/nnano.2015.194} {\bibfield  {journal} {\bibinfo
  {journal} {Nat. Nanotechnol.}\ }\textbf {\bibinfo {volume} {10}},\ \bibinfo
  {pages} {980} (\bibinfo {year} {2015})}\BibitemShut {NoStop}%
\bibitem [{\citenamefont {Cheng}\ \emph {et~al.}(2017)\citenamefont {Cheng},
  \citenamefont {Zhu}, \citenamefont {Han}, \citenamefont {Liu}, \citenamefont
  {Yang}, \citenamefont {Nie}, \citenamefont {Huang}, \citenamefont
  {Shahbazian-Yassar},\ and\ \citenamefont {Mashayek}}]{cheng2017sodium}%
  \BibitemOpen
  \bibfield  {author} {\bibinfo {author} {\bibfnamefont {Y.}~\bibnamefont
  {Cheng}}, \bibinfo {author} {\bibfnamefont {Y.}~\bibnamefont {Zhu}}, \bibinfo
  {author} {\bibfnamefont {Y.}~\bibnamefont {Han}}, \bibinfo {author}
  {\bibfnamefont {Z.}~\bibnamefont {Liu}}, \bibinfo {author} {\bibfnamefont
  {B.}~\bibnamefont {Yang}}, \bibinfo {author} {\bibfnamefont {A.}~\bibnamefont
  {Nie}}, \bibinfo {author} {\bibfnamefont {W.}~\bibnamefont {Huang}}, \bibinfo
  {author} {\bibfnamefont {R.}~\bibnamefont {Shahbazian-Yassar}},\ and\
  \bibinfo {author} {\bibfnamefont {F.}~\bibnamefont {Mashayek}},\ }\bibfield
  {title} {\bibinfo {title} {Sodium-induced reordering of atomic stacks in
  black phosphorus},\ }\href {https://doi.org/10.1021/acs.chemmater.6b05052}
  {\bibfield  {journal} {\bibinfo  {journal} {Chem. Mater.}\ }\textbf {\bibinfo
  {volume} {29}},\ \bibinfo {pages} {1350} (\bibinfo {year}
  {2017})}\BibitemShut {NoStop}%
\bibitem [{\citenamefont {Jeitschko}\ and\ \citenamefont
  {Donohue}(1972)}]{TmP4_CS}%
  \BibitemOpen
  \bibfield  {author} {\bibinfo {author} {\bibfnamefont {W.}~\bibnamefont
  {Jeitschko}}\ and\ \bibinfo {author} {\bibfnamefont {P.~C.}\ \bibnamefont
  {Donohue}},\ }\bibfield  {title} {\bibinfo {title} {{The high pressure
  synthesis, crystal structure, and properties of CrP${\sb 4}$ and MoP${\sb
  4}$}},\ }\href {https://doi.org/10.1107/S0567740872005187} {\bibfield
  {journal} {\bibinfo  {journal} {Acta Crystallogr., Sect. B}\ }\textbf
  {\bibinfo {volume} {28}},\ \bibinfo {pages} {1893} (\bibinfo {year}
  {1972})}\BibitemShut {NoStop}%
\bibitem [{\citenamefont {Mayo}\ \emph {et~al.}(2021)\citenamefont {Mayo},
  \citenamefont {Richards}, \citenamefont {Takahashi},\ and\ \citenamefont
  {Ishiwata}}]{MoP4_HighPressure}%
  \BibitemOpen
  \bibfield  {author} {\bibinfo {author} {\bibfnamefont {A.~H.}\ \bibnamefont
  {Mayo}}, \bibinfo {author} {\bibfnamefont {J.~A.}\ \bibnamefont {Richards}},
  \bibinfo {author} {\bibfnamefont {H.}~\bibnamefont {Takahashi}},\ and\
  \bibinfo {author} {\bibfnamefont {S.}~\bibnamefont {Ishiwata}},\ }\bibfield
  {title} {\bibinfo {title} {High-pressure synthesis of a massive and
  non-symmorphic dirac semimetal candidate $\mathrm{{Mo}P}_4$},\ }\href
  {https://doi.org/10.7566/JPSJ.90.123704} {\bibfield  {journal} {\bibinfo
  {journal} {J. Phys. Soc. Jpn.}\ }\textbf {\bibinfo {volume} {90}},\ \bibinfo
  {pages} {123704} (\bibinfo {year} {2021})}\BibitemShut {NoStop}%
\bibitem [{\citenamefont {Kinomura}\ \emph {et~al.}(1983)\citenamefont
  {Kinomura}, \citenamefont {Terao}, \citenamefont {Kikkawa},\ and\
  \citenamefont {Koizumi}}]{TmP4_CS_WP4}%
  \BibitemOpen
  \bibfield  {author} {\bibinfo {author} {\bibfnamefont {N.}~\bibnamefont
  {Kinomura}}, \bibinfo {author} {\bibfnamefont {K.}~\bibnamefont {Terao}},
  \bibinfo {author} {\bibfnamefont {S.}~\bibnamefont {Kikkawa}},\ and\ \bibinfo
  {author} {\bibfnamefont {M.}~\bibnamefont {Koizumi}},\ }\bibfield  {title}
  {\bibinfo {title} {Preparation and properties of $\mathrm{{WP}}_4$},\ }\href
  {https://doi.org/https://doi.org/10.1016/0022-4596(83)90086-5} {\bibfield
  {journal} {\bibinfo  {journal} {J. Solid State Chem.}\ }\textbf {\bibinfo
  {volume} {48}},\ \bibinfo {pages} {306} (\bibinfo {year} {1983})}\BibitemShut
  {NoStop}%
\bibitem [{\citenamefont {Khan}\ \emph {et~al.}(2020)\citenamefont {Khan},
  \citenamefont {Bu}, \citenamefont {Chai},\ and\ \citenamefont
  {Wang}}]{TmP4_Khan2020}%
  \BibitemOpen
  \bibfield  {author} {\bibinfo {author} {\bibfnamefont {M.~R.}\ \bibnamefont
  {Khan}}, \bibinfo {author} {\bibfnamefont {K.}~\bibnamefont {Bu}}, \bibinfo
  {author} {\bibfnamefont {J.-S.}\ \bibnamefont {Chai}},\ and\ \bibinfo
  {author} {\bibfnamefont {J.-T.}\ \bibnamefont {Wang}},\ }\bibfield  {title}
  {\bibinfo {title} {Novel electronic properties of monoclinic
  $\mathrm{M}\mathrm{P}_4$ ($\mathrm{M} = \mathrm{Cr}, \mathrm{Mo},
  \mathrm{W}$) compounds with or without topological nodal line},\ }\href
  {https://doi.org/10.1038/s41598-020-68349-9} {\bibfield  {journal} {\bibinfo
  {journal} {Sci. Rep.}\ }\textbf {\bibinfo {volume} {10}},\ \bibinfo {pages}
  {11502} (\bibinfo {year} {2020})}\BibitemShut {NoStop}%
\bibitem [{\citenamefont {Hohenberg}\ and\ \citenamefont
  {Kohn}(1964)}]{DFT_1964}%
  \BibitemOpen
  \bibfield  {author} {\bibinfo {author} {\bibfnamefont {P.}~\bibnamefont
  {Hohenberg}}\ and\ \bibinfo {author} {\bibfnamefont {W.}~\bibnamefont
  {Kohn}},\ }\bibfield  {title} {\bibinfo {title} {Inhomogeneous electron
  gas},\ }\href {https://doi.org/10.1103/PhysRev.136.B864} {\bibfield
  {journal} {\bibinfo  {journal} {Phys. Rev.}\ }\textbf {\bibinfo {volume}
  {136}},\ \bibinfo {pages} {B864} (\bibinfo {year} {1964})}\BibitemShut
  {NoStop}%
\bibitem [{\citenamefont {Bl\"ochl}(1994)}]{blochl1994projector}%
  \BibitemOpen
  \bibfield  {author} {\bibinfo {author} {\bibfnamefont {P.~E.}\ \bibnamefont
  {Bl\"ochl}},\ }\bibfield  {title} {\bibinfo {title} {Projector augmented-wave
  method},\ }\href {https://doi.org/10.1103/PhysRevB.50.17953} {\bibfield
  {journal} {\bibinfo  {journal} {Phys. Rev. B}\ }\textbf {\bibinfo {volume}
  {50}},\ \bibinfo {pages} {17953} (\bibinfo {year} {1994})}\BibitemShut
  {NoStop}%
\bibitem [{\citenamefont {Kresse}\ and\ \citenamefont
  {Furthm\"uller}(1996)}]{kresse1996efficient}%
  \BibitemOpen
  \bibfield  {author} {\bibinfo {author} {\bibfnamefont {G.}~\bibnamefont
  {Kresse}}\ and\ \bibinfo {author} {\bibfnamefont {J.}~\bibnamefont
  {Furthm\"uller}},\ }\bibfield  {title} {\bibinfo {title} {Efficient iterative
  schemes for ab initio total-energy calculations using a plane-wave basis
  set},\ }\href {https://doi.org/10.1103/PhysRevB.54.11169} {\bibfield
  {journal} {\bibinfo  {journal} {Phys. Rev. B}\ }\textbf {\bibinfo {volume}
  {54}},\ \bibinfo {pages} {11169} (\bibinfo {year} {1996})}\BibitemShut
  {NoStop}%
\bibitem [{\citenamefont {Kresse}\ and\ \citenamefont
  {Joubert}(1999)}]{kresse1999from}%
  \BibitemOpen
  \bibfield  {author} {\bibinfo {author} {\bibfnamefont {G.}~\bibnamefont
  {Kresse}}\ and\ \bibinfo {author} {\bibfnamefont {D.}~\bibnamefont
  {Joubert}},\ }\bibfield  {title} {\bibinfo {title} {From ultrasoft
  pseudopotentials to the projector augmented-wave method},\ }\href
  {https://doi.org/10.1103/PhysRevB.59.1758} {\bibfield  {journal} {\bibinfo
  {journal} {Phys. Rev. B}\ }\textbf {\bibinfo {volume} {59}},\ \bibinfo
  {pages} {1758} (\bibinfo {year} {1999})}\BibitemShut {NoStop}%
\bibitem [{\citenamefont {Perdew}\ \emph {et~al.}(1996)\citenamefont {Perdew},
  \citenamefont {Burke},\ and\ \citenamefont
  {Ernzerhof}}]{perdew1996generalized}%
  \BibitemOpen
  \bibfield  {author} {\bibinfo {author} {\bibfnamefont {J.~P.}\ \bibnamefont
  {Perdew}}, \bibinfo {author} {\bibfnamefont {K.}~\bibnamefont {Burke}},\ and\
  \bibinfo {author} {\bibfnamefont {M.}~\bibnamefont {Ernzerhof}},\ }\bibfield
  {title} {\bibinfo {title} {Generalized gradient approximation made simple},\
  }\href {https://doi.org/10.1103/PhysRevLett.77.3865} {\bibfield  {journal}
  {\bibinfo  {journal} {Phys. Rev. Lett.}\ }\textbf {\bibinfo {volume} {77}},\
  \bibinfo {pages} {3865} (\bibinfo {year} {1996})}\BibitemShut {NoStop}%
\bibitem [{\citenamefont {Mostofi}\ \emph {et~al.}(2014)\citenamefont
  {Mostofi}, \citenamefont {Yates}, \citenamefont {Pizzi}, \citenamefont {Lee},
  \citenamefont {Souza}, \citenamefont {Vanderbilt},\ and\ \citenamefont
  {Marzari}}]{arash2014an}%
  \BibitemOpen
  \bibfield  {author} {\bibinfo {author} {\bibfnamefont {A.~A.}\ \bibnamefont
  {Mostofi}}, \bibinfo {author} {\bibfnamefont {J.~R.}\ \bibnamefont {Yates}},
  \bibinfo {author} {\bibfnamefont {G.}~\bibnamefont {Pizzi}}, \bibinfo
  {author} {\bibfnamefont {Y.-S.}\ \bibnamefont {Lee}}, \bibinfo {author}
  {\bibfnamefont {I.}~\bibnamefont {Souza}}, \bibinfo {author} {\bibfnamefont
  {D.}~\bibnamefont {Vanderbilt}},\ and\ \bibinfo {author} {\bibfnamefont
  {N.}~\bibnamefont {Marzari}},\ }\bibfield  {title} {\bibinfo {title} {An
  updated version of wannier90: A tool for obtaining maximally-localised
  wannier functions},\ }\href
  {https://doi.org/https://doi.org/10.1016/j.cpc.2014.05.003} {\bibfield
  {journal} {\bibinfo  {journal} {Comput. Phys. Commun.}\ }\textbf {\bibinfo
  {volume} {185}},\ \bibinfo {pages} {2309} (\bibinfo {year}
  {2014})}\BibitemShut {NoStop}%
\bibitem [{\citenamefont {Wu}\ \emph {et~al.}(2018)\citenamefont {Wu},
  \citenamefont {Zhang}, \citenamefont {Song}, \citenamefont {Troyer},\ and\
  \citenamefont {Soluyanov}}]{wu2018wanniertools}%
  \BibitemOpen
  \bibfield  {author} {\bibinfo {author} {\bibfnamefont {Q.}~\bibnamefont
  {Wu}}, \bibinfo {author} {\bibfnamefont {S.}~\bibnamefont {Zhang}}, \bibinfo
  {author} {\bibfnamefont {H.-F.}\ \bibnamefont {Song}}, \bibinfo {author}
  {\bibfnamefont {M.}~\bibnamefont {Troyer}},\ and\ \bibinfo {author}
  {\bibfnamefont {A.~A.}\ \bibnamefont {Soluyanov}},\ }\bibfield  {title}
  {\bibinfo {title} {Wanniertools: An open-source software package for novel
  topological materials},\ }\href
  {https://doi.org/https://doi.org/10.1016/j.cpc.2017.09.033} {\bibfield
  {journal} {\bibinfo  {journal} {Comput. Phys. Commun.}\ }\textbf {\bibinfo
  {volume} {224}},\ \bibinfo {pages} {405} (\bibinfo {year}
  {2018})}\BibitemShut {NoStop}%
\bibitem [{sup()}]{supplementary}%
  \BibitemOpen
  \href@noop {} {\bibinfo {title} {See {S}upplemental {M}aterial}}\BibitemShut
  {NoStop}%
\bibitem [{\citenamefont {Liu}\ \emph {et~al.}(2010)\citenamefont {Liu},
  \citenamefont {Qi}, \citenamefont {Zhang}, \citenamefont {Dai}, \citenamefont
  {Fang},\ and\ \citenamefont {Zhang}}]{Model_TI}%
  \BibitemOpen
  \bibfield  {author} {\bibinfo {author} {\bibfnamefont {C.-X.}\ \bibnamefont
  {Liu}}, \bibinfo {author} {\bibfnamefont {X.-L.}\ \bibnamefont {Qi}},
  \bibinfo {author} {\bibfnamefont {H.}~\bibnamefont {Zhang}}, \bibinfo
  {author} {\bibfnamefont {X.}~\bibnamefont {Dai}}, \bibinfo {author}
  {\bibfnamefont {Z.}~\bibnamefont {Fang}},\ and\ \bibinfo {author}
  {\bibfnamefont {S.-C.}\ \bibnamefont {Zhang}},\ }\bibfield  {title} {\bibinfo
  {title} {Model hamiltonian for topological insulators},\ }\href
  {https://doi.org/10.1103/PhysRevB.82.045122} {\bibfield  {journal} {\bibinfo
  {journal} {Phys. Rev. B}\ }\textbf {\bibinfo {volume} {82}},\ \bibinfo
  {pages} {045122} (\bibinfo {year} {2010})}\BibitemShut {NoStop}%
\bibitem [{\citenamefont {Heyd}\ \emph {et~al.}(2003)\citenamefont {Heyd},
  \citenamefont {Scuseria},\ and\ \citenamefont {Ernzerhof}}]{heyd2006hybrid}%
  \BibitemOpen
  \bibfield  {author} {\bibinfo {author} {\bibfnamefont {J.}~\bibnamefont
  {Heyd}}, \bibinfo {author} {\bibfnamefont {G.~E.}\ \bibnamefont {Scuseria}},\
  and\ \bibinfo {author} {\bibfnamefont {M.}~\bibnamefont {Ernzerhof}},\
  }\bibfield  {title} {\bibinfo {title} {{Hybrid functionals based on a
  screened Coulomb potential}},\ }\href {https://doi.org/10.1063/1.1564060}
  {\bibfield  {journal} {\bibinfo  {journal} {J. Chem. Phys.}\ }\textbf
  {\bibinfo {volume} {118}},\ \bibinfo {pages} {8207} (\bibinfo {year}
  {2003})}\BibitemShut {NoStop}%
\end{thebibliography}%

\end{document}